\documentclass[%
 preprint,
 amsmath,amssymb,
 aps,
 onecolumn,
floatfix,
]{revtex4-2}

\usepackage{graphicx}
\usepackage{dcolumn}
\usepackage{bm}
\usepackage{xcolor}
\usepackage{setspace} 

\usepackage[colorlinks=true,
            linkcolor=blue,
            urlcolor=blue,
            citecolor=blue]{hyperref}

\begin{document}
\sloppy
\section{title}
 \noindent
\textbf{A microscopic Kondo lattice model for the heavy fermion antiferromagnet CeIn$_3$}

\section{Author List}
\noindent
W. Simeth$^1$, Z. Wang$^{2,3,*}$, E. A. Ghioldi$^2$, D. M Fobes$^4$, A.~Podlesnyak$^5$, N.~H.~Sung$^4$, E.~D.~Bauer$^4$, J. Lass$^6$, S. Flury$^{1,7}$, J. Vonka$^1$, D. G. Mazzone$^6$, C. Niedermayer$^6$, Yusuke~Nomura$^8$, Ryotaro Arita$^{8,9}$, C. D. Batista$^{2,10}$, F. Ronning$^4$, M. Janoschek$^{1,4,7}$

\section{Affiliations}
 \noindent
\textit{$^1$Laboratory for Neutron and Muon Instrumentation, Paul Scherrer Institute, Villigen PSI, Switzerland\\
$^2$Department of Physics and Astronomy, The University of Tennessee, Knoxville, TN, 37996, USA\\
$^3$School of Physics and Astronomy, University of Minnesota, Minneapolis, Minnesota 55455, USA\\
$^4$Los Alamos National Laboratory, Los Alamos, NM 87545, USA\\
$^5$Neutron Scattering Division, Oak Ridge National Laboratory, Oak Ridge, Tennessee 37831, USA\\
$^6$Laboratory for Neutron Scattering and Imaging, Paul Scherrer Institute, Villigen PSI, Switzerland\\
$^{7}$Physik-Institut, Universit\"{a}t Z\"{u}rich, Winterthurerstrasse 190, CH-8057 Z\"{u}rich, Switzerland\\
$^8$RIKEN Center for Emergent Matter Science, Wako, Saitama 351-0198, Japan\\
$^9$Department of Applied Physics, The University of Tokyo, Hongo, Bunkyo-ku, Tokyo 113-8656, Japan\\
$^{10}$Quantum Condensed Matter Division and Shull-Wollan Center, Oak Ridge National Laboratory, Oak Ridge, TN, 37831, USA\\
$^*$Present address: Center for Correlated Matter and School of Physics, Zhejiang University, Hangzhou 310058, China}
 
\section{Abstract}
\textbf{Electrons at the border of localization generate exotic states of matter across all classes of strongly correlated electron materials and many other quantum materials with emergent functionality. Heavy electron metals are a model example, in which magnetic interactions arise from the opposing limits of localized and itinerant electrons. This remarkable duality is intimately related to the emergence of a plethora of novel quantum matter states such as unconventional superconductivity, electronic-nematic states, hidden order and most recently topological states of matter such as topological Kondo insulators and Kondo semimetals and putative chiral superconductors. The outstanding challenge is that the archetypal Kondo lattice model that captures the underlying electronic dichotomy is notoriously difficult to solve for real materials. Here we show, using the prototypical strongly-correlated antiferromagnet CeIn$_3$, that a multi-orbital periodic Anderson model embedded with input from ab initio bandstructure calculations can be reduced to a simple Kondo-Heisenberg model, which captures the magnetic interactions quantitatively. We validate this tractable Hamiltonian via high-resolution neutron spectroscopy that reproduces accurately the magnetic soft modes in CeIn$_3$, which are believed to mediate unconventional superconductivity. Our study paves the way for a quantitative understanding of metallic quantum states such as unconventional superconductivity.}
\newline

\section{Introduction}
The Kondo lattice model has been key in qualitatively demonstrating how a myriad of correlated quantum matter states emerge~\cite{2009_Pfleiderer_RevModPhys, 2017_Ronning_Nature,2020_Seo_PhysRevX, 2011_Mydosh_RevModPhys, 2020_Pirie_NatPhys, Lai2018, 2020_Jiao_Nature, Paschen2021} from the interplay of local and itinerant electrons~\cite{2004_Kotliar, 2005_Dagotto, Paschen2021}. Beyond the strongly correlated electron materials for which this archetypal model was conceived, it applies to a growing list of novel quantum systems with potential for applications including the electronic transport through quantum dots~\cite{Cronenwett98}, voltage-tunable magnetic moments in graphene~\cite{Jiang2018}, magnetism in twisted-bilayer graphene~\cite{Javad2021} and in two-dimensional organometallic materials~\cite{Tuerhong2018}, the electronic structure in layered narrow-electronic-band materials~\cite{Shen2022}, electronic resonances of Kagome metals~\cite{Zhang2020}, metallic spin liquid states~\cite{Nakatsuji2006, Tokiwa2014, Tokiwa2015} that may even be of chiral character~\cite{Machida2010}, skyrmions in centrosymmetric magnets~\cite{2019_Kurumaji_Science, 2020_Khanh_NatNanotechnolb} and fully tunable electronic quasiparticles in semiconductor moiré materials~\cite{Zhao2023}. Further, Kondo lattice models have been used to study flat-band materials \cite{2021_Kumar_PhysRevResearch} and  predict novel topological states such as topological superconductivity~\cite{Seifert2018} and quantum spin liquid states~\cite{Hsieh2017}, including the highly sought-after fractional quasiparticles~\cite{Han2022}.  Despite this continued relevance, quantitative predictions for real materials based on Kondo lattice models remain a formidable computational hurdle.

Metals containing cerium are firmly established model systems for the interplay between itinerant and localized electronic degrees of freedom  {\color{black}and are ideal candidates to make progress on this issue}. A prototypical case is CeIn$_3$, whose phase diagram as a function of temperature $T$ and hydrostatic pressure $p$ is shown in Fig.~\ref{fig:1}\textbf{a}~\cite{2001_Knebel_PhysRevB}. The formation of well-localized magnetic moments occurs due to Ce $4f$ orbitals that are buried close to the nuclei. A weak hybridization with conduction electron bands leads to a long-range magnetic exchange interaction between the moments known as the Ruderman-Kittel-Kasuya-Yosida (RKKY) interaction~\cite{Ruderman54,Kasuya56,Yoshida57} (see Fig.~\ref{fig:1}\textbf{b}). By increasing the strength of the hybridization, one can screen the magnetic moment through the Kondo effect, leading to a strongly renormalized electronic density of states near the Fermi energy. This heavy Fermi-liquid state borders the magnetically ordered state, from which it is separated by a magnetic quantum phase transition (QPT) that can be accessed via an external control parameter (here: pressure). Interestingly, novel quantum states (here: superconductivity~\cite{1998_Mathur_Nature}) emerge generically in the vicinity of the QPT suggesting that they are mediated by the associated magnetic quantum critical fluctuations~\cite{Paschen2021}. To understand this emergence near QPTs in a quantitative manner requires a materials-specific microscopic model that incorporates relevant interactions to account for the magnetically ordered state and the resulting fluctuations. 

Both for experiment and theory, the challenge in understanding real materials is the extreme energy resolution ($\sim$meV) required to capture the inherently small energy scales that emerge in the renormalized electronic state (see Fig.~\ref{fig:2}\textbf{a}). On the non-magnetic side of the QPT, a pioneering time-of-flight (TOF) neutron spectroscopy study{~\cite{2018_Goremychkin_Science}, complemented by a subsequent resonant inelastic X-ray (RIXS) scattering measurement~\cite{2022_Rahn}, on a selected material with a relatively large Kondo interaction ($\approx60$\,meV) has only recently achieved quantitative agreement with dynamical mean field theory (DMFT). In contrast, for the magnetic (or superconducting) state of interest here, the relevant energy scale is typically of the order of meV, making this a formidable issue. Materials-specific theoretical investigations of emergent phenomena in $f$-electron materials \cite{ZwicknaglAIP1992, IkedaCeCoIn5PRB2014} are often limited by the difficulty of validating low-energy effective models derived from complex high-energy input. Consequently, studies of such emergent states of matter are generally restricted to oversimplified models of real materials. Experimentally,  in addition to the demand for energy resolution, the pivotal requirement is state-of-the-art momentum-transfer resolution. Notably, the long-range nature of RKKY exchange in real space (Fig.~\ref{fig:1}\textbf{b}) universally results in extremely sharp magnetic excitations in momentum space, as we elucidate below.

Here we illustrate an approach to reduce a multi-orbital periodic Anderson model (MO-PAM) imbued with materials-specific hopping parameters to a minimal Kondo-Heisenberg lattice model that can form the basis for understanding emergent states of matter in various Kondo lattice materials (Fig.~\ref{fig:2}\textbf{a}). Using controlled fourth-order perturbation theory and the high combined momentum-transfer and energy resolution of the latest-generation TOF spectrometers, we resolve the full magnetic interaction over the entire range of relevant length scales. We thereby demonstrate that we can determine accurately the emergent energy scales four orders of magnitude smaller than the bare parameters of the initial MO-PAM. In doing so, we critically evaluate the assumptions made to date about heavy-fermion materials.

\section{Results}

Due to a simple cubic structure that facilitates numerical calculations and its characteristic Doniach phase diagram as a function of pressure $p$~(Fig.~\ref{fig:1}\textbf{a}), CeIn$_3$ is ideally suited to revisit the role of magnetic interactions on a Kondo lattice. As the MO-PAM was originally conceived to explain the localization of shielded electron shells \cite{1961_Anderson_PhysRev} and offers a realistic theoretical treatment of coupled charge and spin degrees of freedoms in correlated metals, we begin with a 25-orbital PAM for CeIn$_3$, 
\begin{equation}
\mathcal{H}_{\textrm{MO-PAM}} = \sum_{\bf{k},s} \epsilon_{\bf{k},s} c^{\dagger}_{\bf{k},s}c_{\bf{k},s} + \sum_{\substack{ i,m, m' \\ \sigma, \sigma'}}  h^{\rm SI}_{m^{\prime} \sigma^{\prime} ; m \sigma}
f_{i,m^{\prime} \sigma^{\prime}}^{\dagger} f_{i,m \sigma} 
+   
\sum_{\bf{k}, s, m, \sigma} V_{\bf{k},m \sigma s}\left( f^{\dagger}_{\bf{k}, m\sigma} c_{\bf{k},s} + h.c. \right),
\label{eq:AndersonModel}
\end{equation}
where $c^{\dagger}_{\bf{k},s}$ ($c_{\bf{k},s}$) is the creation (annihilation) operator of band electrons  with wavevector $\bf{k}$ and band index $s$, which includes the spin index.  $f^{\dagger}_{i,m\sigma}$ ($f_{i,m\sigma}$) denotes the creation (annihilation) operator of an $f$-electron state with $l_z=m$ ($-3 \leq m \leq 3$) and spin $\sigma$ on lattice site $i$. The matrix elements $h^{\rm SI}_{m^{\prime} \sigma^{\prime} ; m \sigma}$ of the single-ion $f$-electron Hamiltonian include  the crystal-field coefficients and intra-atomic spin-orbit coupling (see Methods for details). For CeIn$_3$ and related materials, it is well established that the $f^2$ state is energetically considerably less favorable than the $f^0$ excited state~\cite{SundermannJESRP2016, BraicovichPRB2007}. Thus, we assume that the on-site repulsive interaction between $f$-electrons is infinitely large, implying that $f^2$ configurations are excluded from the Hilbert space: $f^{\dagger}_{i,m\sigma} f^{\dagger}_{i,m'\sigma'}=0$. The dispersion of the conduction electrons ($\epsilon_{\bf{k},s}$) and the hybridization between $f$ and conduction electron states ($V_{\bf{k},m \sigma s}$) are obtained from a tight-binding fit to the ab initio bandstructure calculation (see Methods and Supplementary Information for details). We find it important to include 18 conduction electron orbitals per spin (9 In-$p$, 3 In-$s$, 5 Ce-$d$ and 1 Ce-$s$) to account fully for the electronic structure of the conduction bands near the Fermi level $E_F$, and to obtain well localized $f$-orbitals. The spin-orbit coupling and crystal fields of CeIn$_3$ allow that, out of the 14 $f$ states, only the $\Gamma_7$ ground state doublet needs to be included, which is separated by 12 meV from the $\Gamma_8$ quartet excited state~\cite{2003_Knafo_JPhysCondensMatter}. Hence, the energy of the $\Gamma_7$ state ($\epsilon^{f}_{\Gamma_7}$) is the one remaining free parameter in our model. We further demonstrate in the Supplementary Information that our ab initio band structure calculation is in agreement with the electronic structure determined via angle-resolved photo emission spectroscopy (ARPES) experiments~\cite{2016_Zhang_SciRepa}.

As the hybridization between the conduction electrons and the $f$-electrons is small, we can use degenerate perturbation theory (see Supplementary Information for details) to derive an effective Kondo-Heisenberg Hamiltonian,
\begin{equation}
{\mathcal{H}}_{\mathrm{KH}}= \sideset{}{'}\sum_{\bf{k}, s}  \epsilon_{\bf{k},s} {c}_{\bf{k},s}^\dagger {c}_{\bf{k},s} +\sideset{}{'}\sum_{\substack{i,\bf{k}, \bf{k}^\prime,   \\
\sigma, \sigma^{\prime}, s, s^{\prime} }}  {J}_{i, \sigma \sigma^{\prime}}^{\bf{k} s, \bf{k}^\prime s^{\prime}} \tilde{f}_{i, \sigma}^{\dagger} \tilde{f}_{i, \sigma^{\prime}} {c}_{\bf{k}^\prime, s^{\prime}}^{\dagger} {c}_{\bf{k},s}
+ \sum_{\bf{q}, \mu, \nu} { I}_{\bf{q}}^{\mu\nu} {S}^{\mu}_{\bf{q}}  {S}^{\nu}_{-\bf{q}}, 
\label{eq:KH}
\end{equation}
where $\tilde{f}_{i, \sigma}^{\dagger}$ creates an $f$-electron in the lowest energy $\Gamma_7$ doublet state ($\sigma=\{\uparrow,\downarrow\}$) of site $i$, ${S}^{\nu}_{\bf{q}}$ is the Fourier transform of the effective spin-$1/2$ operator $S^{\nu}_i$ (see Methods). The Kondo coupling is ${J}_{i, \sigma \sigma^{\prime}}^{\bf{k} s, \bf{k}^\prime s^{\prime}}=\frac{1}{N} e^{i(\bf{k}-\bf{k}^\prime) \cdot \bf{r}_{i}} \tilde{V}_{\bf{k}, \sigma s} \tilde{V}_{\bf{k}^\prime, \sigma^{\prime} s^{\prime}}^{*} /(E_F-\varepsilon^{f}_{\Gamma_7})$ where $\tilde{V}_{\bf{k},\sigma s}$ is the hybridization projected to the $\Gamma_7$ doublet. ${\mathcal{H}}_{\mathrm{KH}}$ contains effectively only the two conduction-electron bands that are close to the Fermi energy (the sum $\sum'$ is restricted to band states within a cut-off $\Lambda =0.5$~eV: $|\epsilon_{\bf{k}, s}- E_F| \leq \Lambda$ and  $|\epsilon_{\bf{k}^\prime, s^\prime}- E_F| \leq \Lambda$; cf. top of Fig.~\ref{fig:2}\textbf{b}), as opposed to the 18 conduction bands in the MO-PAM. The remaining bands have been integrated out by including all fourth-order particle-hole and particle-particle  processes that give rise to two types of magnetic interactions between the $f$-moments in $I^{\mu \nu}_{\bf{q}}$ (last term of Eq.~\ref{eq:KH}). Notably, in addition to short-ranged superexchange involving particle-hole and particle-particle processes with excited states outside the cut-off $\Lambda$ ($I_{\bf{q}}^{\rm SE}$), long-ranged interactions arise from fourth order particle-particle processes with excited states inside the cut-off ($\hat I_{\bf{q}}^{(\text{pp})}$) (cf. Fig.~\ref{fig:2}). This derivation provides the microscopic justification for including a Heisenberg term in model studies of the Kondo lattice (e.g. \cite{Si2001,SenthilPRL2003}). As we shall see below, it also plays an important role in understanding the magnetically ordered state of CeIn$_3$. Importantly, this minimal Kondo-Heisenberg model still contains the materials-specific information through the dispersion relation of the conduction electrons and the Heisenberg exchange coupling. 
 
To validate this new minimal Kondo-Heisenberg model for CeIn$_3$ against experiments, and to illustrate the importance of the different contributions to magnetic interactions identified above, we further derive the RKKY Hamiltonian from the Kondo lattice term (first two terms of Eq.~\eqref{eq:KH}) via an additional Schrieffer-Wolf transformation (see Supplementary Information)~\cite{Fazekas_book,Yamada2019_RKKY}.
The resulting effective spin Hamiltonian is
\begin{equation}
{\mathcal{H}}_{\rm spin} = \mathcal{H}_\text{RKKY} 
+ \sum_{\bf{q}, \mu, \nu} { I}_{\bf{q}}^{\mu\nu} {S}^{\mu}_{\bf{q}}  {S}^{\nu}_{-\bf{q}} = \sum_{\bf{q}, \mu, \nu} {\tilde I}_{\bf{q}}^{\mu\nu} {S}^{\mu}_{\bf{q}}  {S}^{\nu}_{-\bf{q}}. \label{eq:spin}
\end{equation}
The effective exchange interaction ${\tilde I}_{\bf{q}}$ is then a sum of the RKKY interaction $I_{\bf{q}}^{\rm RKKY}$, the superexchange $I_{\bf{q}}^{\rm SE}$ and the particle-particle contribution $\hat I_{\bf{q}}^{(\text{pp})}$. It turns out to be practically isotropic, ${\tilde I}^{\mu \nu}_{\bf{q}} \approx \delta_{\mu \nu} {\tilde I}_{\bf{q}}$ , as a consequence of the weak influence of the spin-orbit interaction on the $f-c$ hybridization amplitudes and the suppression of the $f^2$ magnetic virtual states. In Fig.~\ref{fig:2}\textbf{b}, we show all resulting contributions along the path R$\Gamma$XM$\Gamma$. The inset on the upper right corner of Fig.~\ref{fig:2}\textbf{b} shows the position of these high symmetry points in the Brillouin zone that define this path. For comparison with experiment, we note that the antiferromagnetic order of CeIn$_3$ below a N\'eel temperature $T_N=10\,\mathrm{K}$ is characterized by a magnetic propagation vector $\bf{q}_\text{AFM}=\left(\frac{1}{2},\frac{1}{2},\frac{1}{2}\right)$~\cite{1980_Lawrence_PhysRevB} corresponding to the R point (cf. Figs.~\ref{fig:1} and ~\ref{fig:2}). Crucially, as shown in Fig.~\ref{fig:2}\textbf{b}, the RKKY interaction (blue line) typically thought to mediate magnetic order in Kondo lattice materials, has a global minimum near $\Gamma=(0,0,0)$ and only a local minimum at R, demonstrating that it is not responsible for the onset of the observed order. Instead, the superexchange $I_{\bf{q}}^{\rm SE}$ (purple line) exhibits a global minimum at the R point. Finally, the particle-particle contribution near $E_F$ ($\hat I_{\bf{q}}^{(\text{pp})}$, pink line) is mostly flat, but is characterized by a pronounced, cusplike maximum at $\Gamma$. It is then essential to note that the global minimum of the net interaction ${\tilde I}_{\bf{q}}$ (bold cyan line) at the ordering wavevector R arises from the combination of two effects: a compensation between low-energy particle-particle and particle-hole contributions around the $\Gamma$ point and the  short-range antiferromagnetic superexchange interaction generated by the high-energy processes in both channels.

To go beyond the magnetic ground state and to test the theoretically calculated exchange interaction ${\tilde I}_{\bf{q}}$ quantitatively against experiment, we have carried out high-resolution neutron spectroscopy to measure the dispersion of the magnons in CeIn$_3$, which is determined by the magnetic interactions derived above. Notably, the dispersion is given by $E_{\bf{q}}  = S\sqrt{\left({\tilde I}_{\bf{q}_\text{AFM}}-{\tilde I}_{\bf{q}}\right) \left({\tilde I}_{\bf{q}_\text{AFM}}- {\tilde I}_{\bf{q}_\text{AFM}+\bf{q}}\right)}$ with $S=1/2$ and is shown as the solid blue line in Fig.~\ref{fig:3}\textbf{a}, which presents an overview of our results at $T=1.8\,$K. The key signature predicted by our model is a extremely dispersive magnon around the R point arising due to the steep local minimum of the long-ranged RKKY contribution; this was not identified in a previous study with modest resolution by Knafo et al., who instead reported a large magnon gap of more than 1 meV~\cite{2003_Knafo_JPhysCondensMatter}. In contrast, the magnon dispersion determined from our neutron spectroscopy data as well as the observed magnetic intensity are in quantitative agreement with the calculated dispersion and intensity, respectively (see Fig.~\ref{fig:3}). Here the neutron intensities in Fig.~\ref{fig:3} are expressed as the dynamic magnetic susceptibility $\tilde{\chi}''(\bf{Q}, E)$ and were converted to absolute units of $\mu_{\textrm{B}}^2$meV$^{-1}$ for comparison to theory. In particular, the energy and momentum transfer cuts through the magnon spectrum shown in Figs.~\ref{fig:3}\textbf{c} and \textbf{d}, respectively, demonstrate the excellent agreement between experiment and theory. Figure~\ref{fig:4} showcases the steep magnon dispersion in a small region around the R point, which could only be uncovered by the most modern spectrometers (see Methods). Here panels \textbf{a1}, \textbf{b1}, and \textbf{c1} show slices through the R point along the three cubic high-symmetry directions and confirm that the magnon gap is either absent or substantially smaller than the experimental resolution ($\Delta E$~$=$~106~$\mu$eV) as we show in detail in the Supplementary Information. A single fit parameter, $\epsilon_{\Gamma_7}^f=12.009\,\mathrm{eV}$, reproduces the experimental dispersion both with regard to the bandwidth and the magnon velocity. This showcases that our microscopic model,  which starts with an ab initio bandstructure calculation with energy scales of 10 eV, is able to predict magnetic interactions on the order of meV. In addition to being quantitatively accurate, our calculations are robust against small changes of the chemical potential of the order of 10 meV (resolution of our band structure calculation), which modify the slope and bandwidth of the magnon dispersion by less than 1

Finally, we consider the role of short-range Heisenberg superexchange. Although the long-range RKKY interaction is widely credited with mediating magnetic order in Kondo lattice materials~\cite{Paschen2021}, short-range superexchange interactions are commonly employed to fit the observed magnon dispersion~\cite{1987_Broholm_PhysRevLett,vanDijkPRB2000, 2008_Fak_PhysRevBa,2011_Stockert_NatPhys,1979_Houmann_PhysRevBd,2008_Fak_PhysRevBa, 2014_Das_PhysRevLett} and have been also used for CeIn$_3$~\cite{2003_Knafo_JPhysCondensMatter}. As expected short-range superexchange allows us to model the magnon dispersion at high energy and the zone boundary well (cf. Fig.~\ref{fig:3}). Besides providing a microscopic justification for including short-range superexchange interactions, our model also demonstrates that they are not sufficient to explain the magnons near the magnetic zone center. We quantify this statement via the dimensionless parameter $\eta$ that describes ratio of the magnon velocity $v$ to the magnon bandwidth $W$. The experimental magnon bandwidth is $W_\textrm{exp}=2.75(3)\,\mathrm{meV}$, as determined from the energy cuts in Fig.~\ref{fig:3}\textbf{b}. The experimental magnon velocity, $v_\textrm{exp}=38.6(8)\,\mathrm{meV/r.l.u.}$, was inferred from linear fits to the measured dispersion near to the R point along the different high-symmetry directions and averaged over all directions (black dashed lines in Fig.~\ref{fig:4} and Supplementary Information). The parameter $\eta_\textrm{exp}=2.23(6)$ derived from our experiments compares favorably with $\eta_\textrm{MO-PAM}=2.27(5)$ computed from our model. In contrast, fits of the magnon dispersion with a single exchange constant $J_1$ result in $\eta_{J_1}=0.58$, which stems from a substantial underestimation of the magnon velocity near the R point (cf. light green line in Fig.~\ref{fig:4}). We note that even adding a large number of additional higher-order exchange terms does not allow the large observed magnon velocity to be modelled accurately (see Supplementary Information), as was noted in previous studies~\cite{1979_Houmann_PhysRevBd, 1987_Broholm_PhysRevLett}.

\section{Discussion}

Our study demonstrates that it is now possible to derive a minimal microscopic Kondo lattice model for a real material starting from ab initio electronic bandstructure calculations starting at energy scales of eV. This enables us to reproduce quantitatively the magnetic order and all relevant magnetic interactions in the prototypical Kondo lattice CeIn$_3$ at energy scales 10,000 times smaller. Thus, it provides a tractable path for the accurate calculation of the low-energy spin excitations that arise due to strong electronic correlations, and are believed to mediate the emergence of novel quantum phases~(cf. Fig.~\ref{fig:1}). Due to the long-range nature of the RKKY interaction, these magnetic soft modes are remarkably steep. This is also borne out by neutron scattering studies on further $f$-electron materials, which highlight the broad relevance of our results. Either steep magnon dispersions were observed directly~\cite{1979_Houmann_PhysRevBd, 1987_Broholm_PhysRevLett, 2007_Wiebe_NatPhysa} or this key feature is concealed due to inadequate resolution, resulting in reports of potentially spurious magnon gaps~\cite{OsbornJMMM1987, vanDijkPRB2000, 2014_Das_PhysRevLett} similar to CeIn$_3$~\cite{2003_Knafo_JPhysCondensMatter}. We note that the ability to resolve these steep low-energy spin excitations also ushers in high-resolution TOF spectroscopy as a complementary technique to access the electronic band structure of magnetically ordered heavy-fermion materials.

A further remarkable insight revealed by our calculations is that, in addition to the RKKY interaction,  contributions from the particle-particle channel are equally crucial to quantitative description of the magnetic order and the magnon spectrum. In turn, our approach resolves several puzzles concerning magnetic order in heavy-fermion materials. First, it is consistent with neutron scattering studies of the magnon spectrum of various heavy fermion-materials, where the dispersion towards the zone boundary is well explained using short-range interactions~\cite{1987_Broholm_PhysRevLett,vanDijkPRB2000, 2008_Fak_PhysRevBa,2011_Stockert_NatPhys,1979_Houmann_PhysRevBd,2008_Fak_PhysRevBa, 2014_Das_PhysRevLett}. Second, the presence of short-range interactions highlights why a large collection of heavy fermion materials exhibit commensurate AFM order~\cite{1980_Lawrence_PhysRevB, KawarazakiPRB2000, vanDijkPRB2000, BaoPRB2001, GauthierPRB2017}, even though RKKY interactions arising from generic Fermi surfaces typically favor incommensurate order. Finally, the apparent lack of 4$f$-based metallic ferromagnets is now understood quite simply by the presence of particle-particle interactions in metals, which generically disfavor the ferromagnetic ground state. 

In summary, the combination of our MO-PAM approach with TOF spectroscopy establishes a straightforward recipe to obtain a quantitative, yet relatively simple, effective Kondo-Heisenberg Hamiltonian, which for CeIn$_3$ includes only 2+1 orbitals (two conduction electron bands and one $f$ level). In turn, studying this effective model as a function of pressure has the promising prospect of establishing the emergence of unconventional superconductivity in CeIn$_3$ quantitatively. Notably, the short-range interaction promotes a local-moment magnetic state relative to the bare Kondo temperature, which may fundamentally alter the nature of the magnetic QPT observed under pressure~\cite{SettaiJPSJCeIn3}. Adding the $\Gamma_8$ $f$-state to our calculation should equally allow us to reproduce the magnetic anisotropy emerging as a function of magnetic field~\cite{2017_Moll_NpjQuantumMater}. Considering the ever-increasing computational power available, our approach is equally in reach for more complex materials with lower symmetry and more orbitals, which will allow us to unlock the microscopic understanding of a large number of quantum matter states with functional properties~\cite{Paschen2021}. Further, our discovery that both short- and long-range interactions are key to understanding magnetic order in heavy fermion materials offers a straightforward explanation of their rich magnetic phase diagrams, where changing the balance of interactions by applying external tuning parameters allows us to select distinct magnetic ground states. Similarly, we anticipate that the combination of short- and long-ranged interactions will turn out to be a generic feature of metallic systems whose starting point is the periodic Anderson model. Finally, our study paves the way for ab initio modeling of quantum systems described by Kondo lattices and therefore suggests an avenue beyond a phenomenological description of the ground states of their strongly correlated electron systems.

\clearpage
\section{Methods}

\textbf{Multi-Orbital Periodic Anderson Model}
The input parameters for the multi-orbital periodic Anderson model (MO-PAM) presented in Eq.~\eqref{eq:AndersonModel} specific to CeIn$_3$ were obtained by deriving a tight-binding model based on density functional theory (DFT) that accurately captures the electronic structure of the conduction bands near the Fermi level $E_F$, and yields the hybridization between these bands and the $4f$-orbitals. The underlying DFT band structure calculations were performed using the \textsc{Quantum ESPRESSO} package~\cite{Gianozzi_2017} and  fully relativistic projector augmented-wave (PAW) pseudopotentials with the Perdew-Burke-Ernzerhof (PBE) exchange-correlation functional, which are available in PSlibrary~\cite{Dal_Corso_2014}. A realistic tight-binding Hamiltonian  with 50 Wannier functions (25 orbitals times 2 for spin) was constructed using the Wannier90 package~\cite{Pizzi_2020}. Details for all steps are provided in the Supplementary Material.

The low-temperature magnetic properties of CeIn$_3$ at zero field are dominated by the low-energy Ce 4$f$ $\Gamma_7$ doublet that results from diagonalizing the single-site $f$-electron Hamiltonian [second term of Eq.~\eqref{eq:AndersonModel}],
\begin{equation}
h^{\rm SI}_{m^{\prime} \sigma^{\prime} ; m \sigma}=\epsilon^f  \delta_{\sigma, \sigma'} \delta_{m, m'}+ B_{m,m'} \delta_{\sigma,\sigma'} + \lambda \zeta_{m^{\prime}, \sigma^{\prime};m \sigma},
\end{equation}
which includes the crystal field coefficients $B_{m,m'}$ and the {\it intra-atomic} spin-orbit coupling $\lambda$ ($\zeta_{m' \sigma' ; m \sigma}=\delta_{m',m} \delta_{\sigma',\sigma} m \sigma / 2  +\delta_{m', m+\sigma}\delta_{\sigma', -\sigma} \sqrt{12-m(m+\sigma)} / 2  $).
The resulting $\Gamma_7$ doublet reads
\begin{subequations}
\begin{align}
|\Gamma_7;+\rangle &\equiv |\uparrow \rangle = \sqrt{\frac{1}{6}} \left|j=\frac{5}{2},m_j=\frac{5}{2}\right\rangle - \sqrt{\frac{5}{6}} \left|j=\frac{5}{2},m_j=-\frac{3}{2}\right\rangle,\\
|\Gamma_7;-\rangle &\equiv |\downarrow \rangle = \sqrt{\frac{1}{6}} \left|j=\frac{5}{2},m_j=-\frac{5}{2}\right\rangle - \sqrt{\frac{5}{6}} \left|j=\frac{5}{2},m_j=\frac{3}{2}\right\rangle.
\end{align}
\end{subequations}
By taking the limit of infinite intra-atomic $f$-$f$ Coulomb repulsion which eliminates the $f^2$ configurations, we project $\mathcal{H}_{\text{MO-PAM} }$ into the low-energy subspace generated by the $\Gamma_7$ doublet and obtain the periodic Anderson model 
\begin{equation}
\mathcal{H}_{\rm PAM} = \sum_{\bf{k}, s} \epsilon_{\bf{k}, s} c_{\bf{k}, s}^{\dag} c_{\bf{k}, s}^{} + \epsilon_{\Gamma_7}^{f} \sum_{i,\sigma} \tilde f_{i, \sigma}^{\dag} \tilde f_{i,\sigma}^{} + \sum_{\bf{k}, \sigma,s} \left( \tilde{V}_{\bf{k},\sigma s} \tilde f_{\bf{k},\sigma}^{\dag} c_{\bf{k}, s}  + h.c. \right) ,
\label{PAM2}
\end{equation}
where the constrained operators $\tilde f_{i, \sigma}^{\dag}$ ($\tilde f_{i, \sigma}$) create (annihilate) an $f$-electron in the $\Gamma_7$ doublet with $\sigma=\{\uparrow,\downarrow \}$ and ${\tilde f}^{\dagger}_{i, \sigma} {\tilde f}^{\dagger}_{i, \sigma'}=0$, $\epsilon_{\Gamma_7}^{f}$ is the energy of the $\Gamma_7$ states, and $\tilde{V}_{\bf{k},\sigma s}$ is the hybridization between the $\Gamma_7$ doublet and the conduction electron states ($1 \leq s \leq 36$).

By treating the small hybridization $\tilde{V}_{\bf{k},\sigma s}$ as a perturbation, the periodic Anderson model can be reduced to the effective Kondo-Heisenberg Hamiltonian shown in Eq.~\eqref{eq:KH}. Furthermore, for strongly localized $f$-electrons, the  Kondo lattice model can be further reduced to the RKKY  Hamiltonian via second order degenerate perturbation theory in the Kondo interaction (see Supplementary Information for details). The final effective spin Hamiltonian $\mathcal{H}_{\text{spin}}$ is presented in Eq.~\eqref{eq:spin}, where the pseudo spin-$\frac{1}{2}$ operator is defined as $ \bf{S}_{i}\equiv\frac{1}{2}\tilde f_{i,\alpha}^{\dagger}\bf{\sigma}_{\alpha\beta}\tilde f_{i,\beta}$ ($\sigma^{\nu}$ are the Pauli matrices with $\nu=x,y,z$), and its Fourier transform is defined as
${S}^{\nu}_{\bf{q}}=\frac{1}{N} \sum_{i} e^{i \bf{q} \cdot \bf{r}_i} S_i^\nu$. $N$ is the total number of Ce atoms on the lattice (we take $N \rightarrow \infty$ in the following calculation)

\textbf{Sample Preparation}
To overcome the high absorption of cold neutrons by indium, plate-like single crystalline samples of CeIn$_3$ were grown by the indium self-flux method and polished to a thickness of around 0.7$\,$mm. To maximize the total scattering intensity 24 pieces were carefully coaligned on an aluminum sample-holder using a hydrogen-free adhesive (CYTOP) (see Sec.~II.A of the Supplementary Information).

\textbf{Neutron Spectroscopy}
Inelastic neutron scattering was carried out at the cold neutron chopper spectrometer CNCS at ORNL~\cite{2011_Ehlers_RevSciInstrum}. For the measurements, the crystal array was oriented in such a way that the crystallographic $\left[1\overline{1}0\right]$ axis was vertical. Momentum transfers of neutrons are given in the reference frame of the sample by means of $\bf{Q}=H \bf{b}_1+K \bf{b}_2+L \bf{b}_3$, where $H$, $K$, and $L$ denote the Miller indices and $\bf{b}_\nu=\frac{2\pi}{a} \hat{\bf{a}}_\nu$ ($\nu=1,2,3$) represent primitive translation vectors of the reciprocal cubic lattice ($a=4.689$~\AA). Throughout the manuscript and the SI, when stating components of momentum transfers the reciprocal lattice unit ($1\cdot\mathrm{r.l.u.}:=\frac{2\pi}{a}$) is omitted. To a given momentum transfer $\bf{Q}$ (upper-case letter), the reduced momentum transfer that equals the equivalent reciprocal space position in the cell $0\leq q_\nu<1$, is given by $\bf{q}=q_1 \bf{b}_1+q_2 \bf{b}_2+q_3 \bf{b}_3$ (labelled by a lower-case letter). $\Delta q$ denotes the modulus of $\Delta \bf{q}$. Reciprocal-space distances are also given in units $1\cdot\mathrm{r.l.u.}:=\frac{2\pi}{a}$.

Time-of-flight neutron spectroscopy was performed with two different incident neutron energies. High-resolution experiments with incident neutron energy $E_{i,1}=3.315\,\mathrm{meV}$ ($\Delta E$~$=$~106~$\mu$eV) permitted the study of the steep magnon dispersion in the vicinity of the reciprocal space position $\bf{Q}_0=\left(-\frac{1}{2},-\frac{1}{2},\frac{1}{2}\right)$, which represents the R point. A ``high-energy'' setting with incident neutron energy $E_{i,2}=12\,\mathrm{meV}$ was performed to determine the magnon dispersion across the entire Brillouin zone. Data were collected in terms of a so-called Horace scan~\cite{2016_Ewings_NuclearInstrumentsandMethodsinPhysicsResearchSectionAAcceleratorsSpectrometersDetectorsandAssociatedEquipment}, where the crystal is rotated around the vertical axis. In combination with the CNCS detector, which has large horizontal (from -50 to 140 degrees) and vertical ($\pm$ 16 degrees) coverage, this allows one to obtain four dimensional data sets covering all three momentum transfer directions and energy transfer. Detailed information is provided in Sec.~II of the Supplementary Information.

For Figs.~\ref{fig:3} and~\ref{fig:4} the neutron intensity recorded in our experiments has been expressed as the imaginary part of the dynamic magnetic susceptibility $\tilde{\chi}''(\bf{Q}, E)$ in absolute units of $\mu_{\textrm{B}}^2$meV$^{-1}$ to enable direct comparison with the calculations of the dynamic magnetic susceptibility via the Multi-Orbital Periodic Anderson Model (see above). For this purpose, the recorded neutron intensities have been corrected for neutron absorption, put on an absolute scale via comparison to the well-known incoherent scattering of the sample, and finally background subtracted via data sets obtained below $T_N$(for details of this procedure, we refer to the Supplementary Information). Further, the resulting corrected and normalized magnetic intensity was then corrected by the Bose factor accounting for the thermal population of magnon states, and subsequently divided by the square of the magnetic form-factor for Ce$^{3+}$ as well as the ratio $\left|\bf{K}_f\right|/\left|\bf{K}_i\right|$, whereby $\bf{K}_i$ and $\bf{K}_f$ denote the wavevectors of incident and scattered neutrons, respectively. For further information, we refer to the Supplementary Information. In the following, we additionally detail the integration ranges for the data shown in Figs.~\ref{fig:3} and~\ref{fig:4}. The data in Fig.~\ref{fig:3}\textbf{a} on the paths $\Gamma$R, R$\Gamma$,$\Gamma X$, XM, and M$\Gamma$ were taken from the reciprocal space lines between the $\bf{Q}$-positions $\left(-1,-1,1\right)$ and $\left(-\frac{1}{2},-\frac{1}{2},\frac{1}{2}\right)$, between $\left(-\frac{1}{2},-\frac{1}{2},\frac{1}{2}\right)$ and $\left(0,0,0\right)$, between $\left(1,1,0\right)$ and $\left(1,1,\frac{1}{2}\right)$, between $\left(-1,-\frac{1}{2},1\right)$ and $\left(-\frac{1}{2},-\frac{1}{2},1\right)$, and between $\left(\frac{1}{2},\frac{1}{2},0\right)$ and $\left(1,1,0\right)$, respectively. The intensity was integrated within a distance of $\pm 0.17\,\mathrm{r.l.u.}$ along the two $\bf{Q}$-directions that are perpendicular to the path $\Gamma$XM and a distance of $\pm 0.09\,\mathrm{r.l.u.}$ along the two $\bf{Q}$-directions that are perpendicular to the paths R$\Gamma$ and M$\Gamma$. For the energy cuts shown Fig.~\ref{fig:3}\textbf{b} the intensity was integrated within a distance of $\pm 0.17\,\mathrm{r.l.u.}$ along three different reciprocal space directions. For the reduced momentum transfer cuts and slices shown in  Fig.~\ref{fig:4}, the intensity was integrated within a distance of $0.08\,\mathrm{r.l.u.}$ along the two reciprocal space directions that are perpendicular to the reduced momentum transfer $\bf{q}$. For the cuts in the lower panels, intensity was integrated over energies $\pm0.1\,\mathrm{meV}$. 

We note that to resolve the sharp magnon dispersion high momentum resolution is required in addition to extreme energy resolution. Although triple-axis neutron spectroscopy is often credited with the best combined momentum-energy-transfer resolution, it fails to properly identify the dispersion even when used with the best resolution due to so-called Currat-Axe spurions ~\cite{2002_Shirane_TAS}. To demonstrate this we have carried out additional measurements on the multiplexing triple axis spectrometer CAMEA at Paul Scherrer Institute~\cite{2020_Lass_CAMEA} using the identical sample and incident energies $E_{i}=4.15\,\mathrm{meV}$, $4.8\,\mathrm{meV}$, and $5.5\,\mathrm{meV}$. Details of these measurements and the spurious features close to the zone center are shown in the Supplementary Information. This highlights the importance of modern TOF spectroscopy for the investigation of strongly correlated metals. 

\section{Data Availability}
The datasets generated during and/or analysed during the current study have been deposited on the ONCat platform of Oak Ridge National Laboratory (ORNL) under the following digital object identifier: 10.14461/oncat.data.64d61647fd6850c0afce4da3/1994514 (\url{https://dx.doi.org/10.14461/oncat.data.64d61647fd6850c0afce4da3/1994514}). Currently, the DOI-minting functionality in ONCat is still in beta and behind a firewall and you need to be on ORNL's network to access it. This can be achieved going to  \url{https://analysis.sns.gov/} and registering an account, signing and starting a session. As soon as the DOI-minting functionality is publically available the data will directly available via the link above. The processed neutron spectroscopy data shown in the manuscript are available at the Zenodo database under accession code digital object identifier: 10.5281/zenodo.10146787 (\url{https://zenodo.org/records/10146787}) \cite{simeth_2023_10146787}.

\section{Code Availability}
All code for the absorption correction of the neutron spectroscopy data is available at the Zenodo database under accession code digital object identifier: 10.5281/zenodo.10147126 (\url{https://zenodo.org/records/10147126}) \cite{wolfgangsimeth_2023_10147126}. 

\section{References}
\bibliographystyle{naturemag}

\providecommand{\noopsort}[1]{}\providecommand{\singleletter}[1]{#1}

\newpage

\section{Acknowledgments}
MJ would like to thank Georg Ehlers for useful discussions concerning the setup of the CNCS experiment and acknowledges fruitful discussions with Bruce Normand. Work at Los Alamos National Laboratory was performed under the U.S. DOE, Office of Science, BES project ``Quantum Fluctuations in Narrow Band Systems''. This research used resources at the Spallation Neutron Source, a DOE Office of Science User Facility operated by the Oak Ridge National Laboratory. WS is supported through funding from the European Union’s Horizon 2020 research and innovation programme under the Marie Sklodowska-Curie grant agreement No 884104 (PSI-FELLOW-III-3i). ZW was supported by funding from the Lincoln Chair of Excellence in Physics. During the writing of this paper, ZW was supported by the U.S. Department of Energy through the University of Minnesota Center for Quantum Materials, under Award No. DE-SC-0016371. FR thanks the hospitality of the University of Tokyo. YN and RA acknowledge funding through Grant-in-Aids for Scientific Research (JSPS KAKENHI, Japan) [Grant No. 20K14423 and 21H01041, and 19H05825, respectively] and ``Program for Promoting Researches on the Supercomputer Fugaku'’ (Project ID:hp210163) from MEXT, Japan.

\section{Author contributions}
CDB, FR and MJ conceived the study. YN, RA and FR carried out the electronic band structure calculations. ZW, EAG and CDB derived the multi-orbital periodic Anderson model and Kondo-Heisenberg model. DMF and MJ designed the TOF experiments. WS, DGM and MJ designed the CAMEA experiments. DMF, WS, AP, JL, SF, JV, DGM, CN and MJ carried out the experiments. JV comissioned and set up the low-temperature sample environment for CAMEA. NHS and EDB synthesized and characterized the samples. DMF assembled the single-crystal mosaic. WS and MJ analyzed the data. WS, ZW, CDB, FR and MJ interpreted the results. WS, ZW, CDB, FR and MJ wrote the paper with input from all the authors.

\section{Competing Interest Statement}
The Authors declare no competing interests. Correspondence and requests for materials should be addressed to M. Janoschek (marc.janoschek@psi.ch).

\newpage
\section{Figure Captions}

\begin{figure*}[h!t]
\includegraphics[width=1\columnwidth]{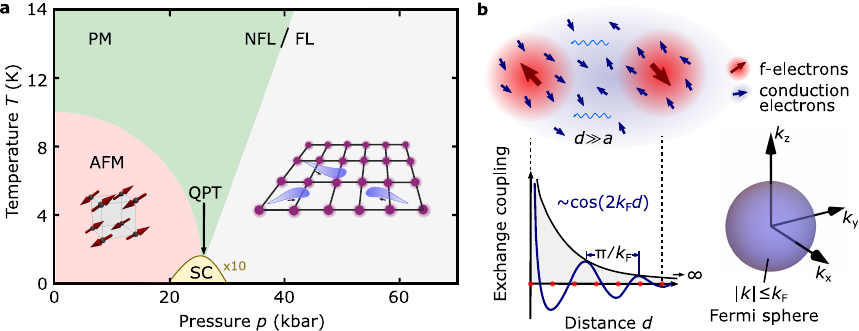}
\caption{\label{fig:1}\textbf{Magnetic interactions and emergent quantum states in a prototypical Kondo lattice}. \textbf{a} The magnetic phase diagram of the heavy electron material CeIn$_3$ as a function of pressure $p$ exemplary of many Kondo lattice materials. The Ruderman-Kittel-Kasuya-Yosida (RKKY) interaction arises when local cerium $4f$ moments polarize the surrounding conduction electron spins, which, in turn, interact with local moments on neighboring Ce sites (see panel \textbf{b}). At ambient pressure  this results in antiferromagnetic (AFM) order. However, increasing pressure increases the overlap of neighboring $4f$ electron wave functions, and, in turn, the hybridization with the conduction electrons. The Kondo effect then leads to the quenching of the local Ce moments, and the formation of strongly-renormalized non-magnetic heavy Fermi liquid (FL). The AFM and FL state are separated by a magnetic quantum phase transition (QPT). The underlying magnetic interactions are therefore not only responsible for the AFM order at ambient pressure, but the associated magnetic quantum fluctuations at the QPT drive the emergence of novel quantum states, in this case, of unconventional superconductivity (SC) and lead to non-Fermi liquid (NFL) behavior. \textbf{b}. Because the RKKY interaction is mediated via conduction electrons (top panel) it is extremely long-range (bottom panel), compared to the lattice parameter $a$. It is also oscillatory in nature as we illustrate for a spherical Fermi-surface with radius $k_F$ (right panel), which results in a period $\lambda \approx \pi/k_F$ in real space (bottom panel). However, as we demonstrate in Fig.~\ref{fig:2}, magnetic order on a Kondo lattice is not only mediated by RKKY interactions but additionally requires short-range superexchange and fourth-order particle-particle interactions.
} 
\end{figure*}

\begin{figure*}[h!t]
\includegraphics[width=1\columnwidth]{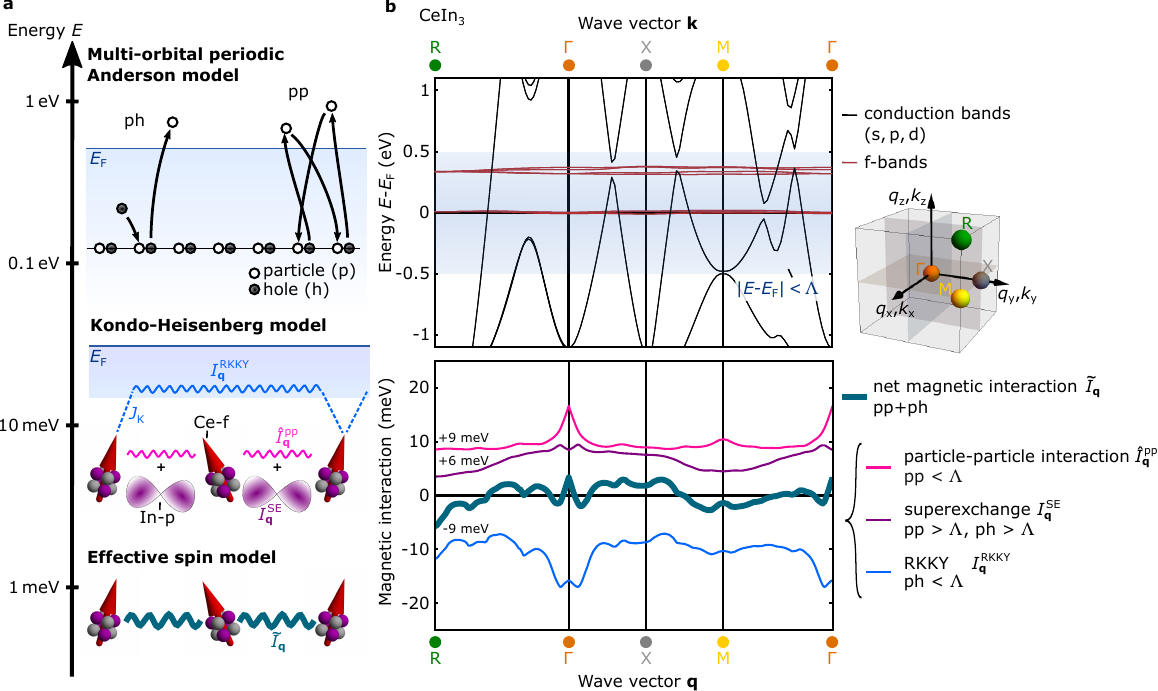}
\caption{\label{fig:2}\textbf{Derivation of the Kondo-Heisenberg model}. \textbf{a} Materials-specific microscopic parameters are contained in the multi-orbital periodic Anderson lattice model (MO-PAM), where $f$-electrons are well localized due to a strong Coulomb repulsion. These states, however, still fluctuate through hybridization with the conduction electron bands. By integrating out the high-energy conduction band states, one arrives at a minimal Kondo + Heisenberg lattice model (KHM). The Kondo interaction contains the spin exchange interaction with the conduction electrons (blue dashed line), while the Heisenberg interaction is a combination of short-range interactions arising from completely filled or completely unoccupied conduction bands (purple $p$-like orbital between the local moments) and a long range interaction due to particle-particle processes not captured through the Kondo exchange (right process in the MO-PAM sketch, and pink wavy line in the KHM sketch). This Hamiltonian encapsulates the physics of heavy-fermion materials such as CeIn$_3$. Finally, the low-energy conduction electrons can be integrated out to yield an effective spin Hamiltonian between local $f$-moments, which contains both the long-range RKKY interaction from the Kondo exchange (blue wavy line), and the other contributions explained above. \textbf{b} Results derived for CeIn$_3$ using this approach are shown. The upper panel shows the electronic bandstructure that serves as input for the MO-PAM along the path R$\Gamma$XM$\Gamma$. The inset on the right illustrates the position of these high symmetry points in the Brillouin zone. The cut-off $\Lambda =0.5\,$eV around the Fermi-energy $E_F$ allows short-range superexchange and long-ranged interactions to be seperated (see text for details). The lower panel shows the various contributions to the net magnetic interaction $\tilde{I}_{\bf{q}}$ in CeIn$_3$ shifted by the indicated energies for visibility. The color of each contribution denotes the corresponding process of the same color in \textbf{a}. 
}
\end{figure*}

\begin{figure*}[h!t]
\includegraphics[width=0.85\columnwidth]{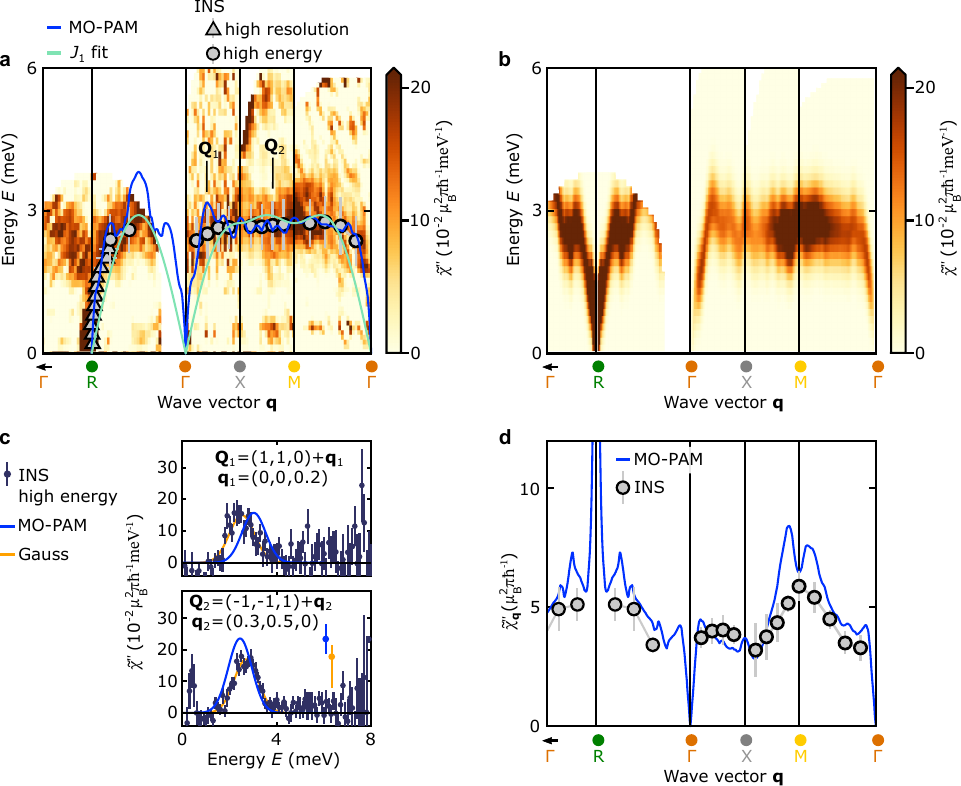}
\caption{\label{fig:3}\textbf{Calculated and measured magnon dispersion and dynamic magnetic susceptibility in the antiferromagnetic state of CeIn$_3$.} Panels \textbf{a} and \textbf{b} show the imaginary part of the dynamic magnetic susceptibility $\tilde{\chi}''(\bf{Q}, E)$ obtained by our experiments and calculated via the MO-PAM model, respectively. Solid lines denote the magnon dispersion on the path $\gamma=$R$\Gamma$XM$\Gamma$ as calculated based on the effective spin interaction $\tilde{I}_{\bf{q}}$ [Eq.~\eqref{eq:spin}] derived from the MO-PAM (blue solid line, see text for details), a Heisenberg model with a single nearest-neighbor exchange constant $J_1$ (light green line), and the dispersion inferred from our neutron spectroscopy measurements (symbols). We refer to the inset on the upper right corner of Fig.~\ref{fig:2}\textbf{b} for the position of the high symmetry positions $R$, $\Gamma$, X, M, and $\Gamma$ in the Brillouin zone. Experimental data points and error bars represent the location and standard deviation of the Gaussian peaks inferred from neutron spectroscopy data sets with incident neutron energies 3.315 meV (high-resolution, triangles) and 12 meV (high-energy, circles), by means of constant-energy and constant-momentum transfer cuts, respectively. All cuts investigated and the corresponding fits are shown in the Supplementary Information. The color-scale provides $\tilde{\chi}''(\bf{Q}, E)$ in absolute units of $\mu_{\textrm{B}}^2$meV$^{-1}$ along the path $\gamma$, as inferred from neutron spectroscopy data taken in the high-energy setting. For details see Methods. \textbf{c} Typical constant-momentum-transfer cuts through $\tilde{\chi}''(\bf{Q},E)$ as illustrated for the two positions $\bf{Q}_{1}$ and $\bf{Q}_{2}$ on the paths X$\Gamma$ and XM, respectively. Square symbols denote high-intensity neutron spectroscopy data and the blue solid lines MO-PAM calculations. Error bars represent the statistical error. The orange line corresponds to a Gaussian fit to the INS data. The orange and blue bars denote the systematic error in the conversion of the neutron intensity to absolute units and uncertainties in the calculation of the MO-PAM intensities, respectively (see Supplementary Information for details). \textbf{d} The integral of $\tilde{\chi}''(\bf{Q},E)$ over energy, denoted $\tilde{\chi}''_{\bf{Q}}$. Circle symbols and the blue solid line correspond to $\tilde{\chi}''_{\bf{Q}}$ from high-intensity neutron data and MO-PAM calculations, respectively. Error bars represent the statistical error.}
\end{figure*}


\begin{figure*}[h!t]
\includegraphics[width=1\columnwidth]{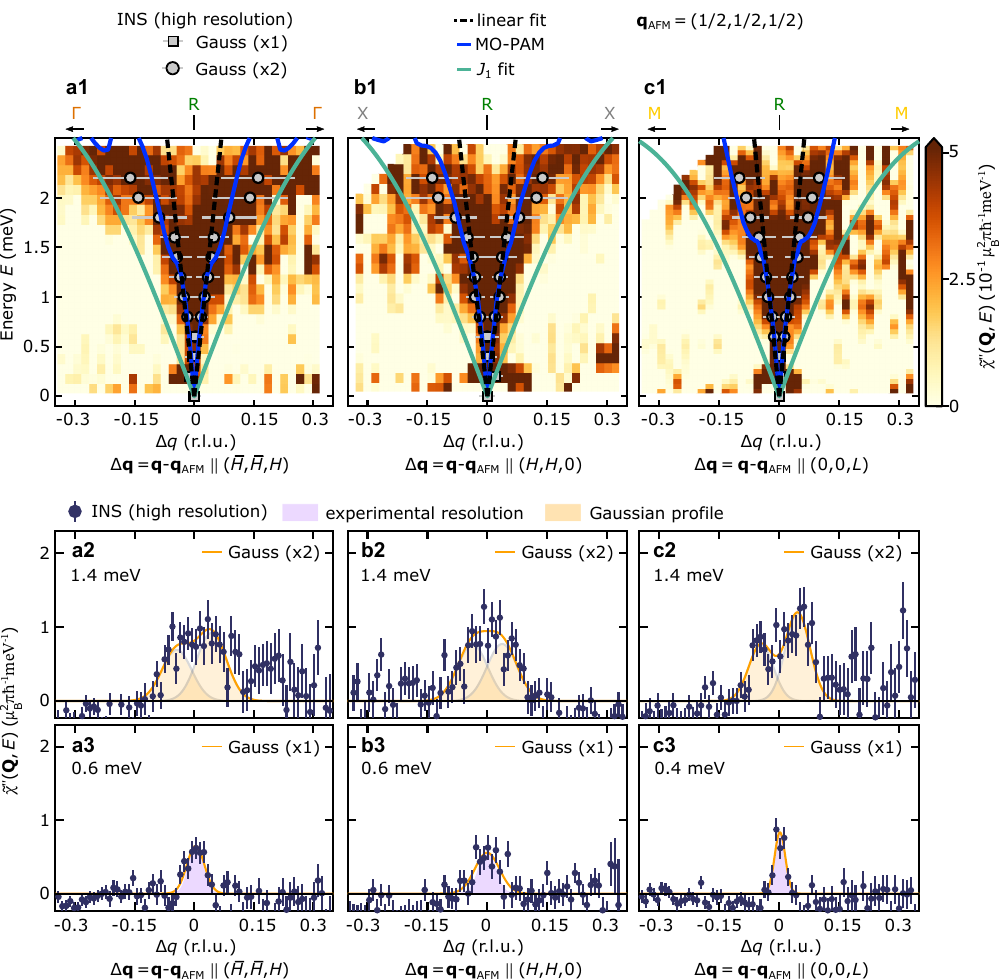}
\caption{\label{fig:4}\textbf{Signature of long-range RKKY interactions in CeIn$_3$.} \textbf{a1},\textbf{b1},\textbf{c1} Magnon dispersion near the magnetic zone center R along the respective lines R$\Gamma$, RX, and RM. The color-coding reflects the imaginary part of the dynamic susceptibility, $\tilde{\chi}''(\bf{Q}, E)$, recorded with incident neutron energy $E_i = 3.315\,\mathrm{meV}$ (see Methods for details). Squares and circles correspond to the locations of Gaussian profiles inferred from constant-energy cuts with one and two Gaussian profiles, respectively. All cuts investigated and the corresponding fits are shown in the Supplementary Information. The solid blue and light green lines denote the magnon dispersions calculated from our microscopic model and a fit of a short-range Heisenberg model with a single nearest-neighbor exchange constant $J_1$ to our data, respectively. The dashed black line is a linear fit of the magnon dispersion close to the zone center used to extract the magnon velocity. \textbf{a2},\textbf{b2},\textbf{c2} At higher energies, where the splitting is larger than the FWHM of the experimental resolution, the profiles of constant-energy cuts were fitted by two Gaussian profiles. \textbf{a3},\textbf{b3},\textbf{c3} At energies below $0.6\,\mathrm{meV}$ for the R$\Gamma$ and RX direction and below $0.4\,\mathrm{meV}$ for the RM direction, the splitting is restricted by the experimental resolution as indicated by the shaded profile. Error bars in panels \textbf{a2},\textbf{b2},\textbf{c2} and \textbf{a3},\textbf{b3},\textbf{c3} represent the statistical error.}
\end{figure*}

\end{document}



\title{Supplementary information for\\A microscopic Kondo lattice model for the heavy fermion antiferromagnet CeIn$_3$}

\begin{abstract}
This supplementary information contains details on the theoretical and experimental methods of our study. First, we provide the details for our electronic structure calculations for CeIn$_3$ performed via density functional theory alongside details for our tight-binding model. Subsequently, we elaborate on the multi-orbital periodic Anderson model, which is the starting point for our microscopic model, and show how we derive the effective low-energy Hamiltonians. In the second part we provide details on the experiments. The experimental methods part starts with an account on the sample preparation and quality. After a description of the instrumental setup used for the inelastic neutron scattering experiments, we discuss the data analysis used to determine the dispersion of magnetic excitations. Subsequently, the experimentally determined magnon dispersion for CeIn$_3$ is compared with the results from the microscopic model.  Finally, the electronic band structure obtained via density functional theory is compared to electronic band structure of CeIn$_3$ obtained via angle-resolved photo emission spectroscopy (ARPES) experiments.
\end{abstract}

\author{W. Simeth$^1$}

\author{Z. Wang$^{2,3,*}$}

\author{E. A. Ghioldi$^2$}

\author{D. M Fobes$^4$}

\author{A.~Podlesnyak$^5$}

\author{N.~H.~Sung$^4$}

\author{E.~D.~Bauer$^4$}

\author{J. Lass$^6$}

\author{S. Flury$^{1,7}$}

\author{J. Vonka$^1$}

\author{D. G. Mazzone$^6$}

\author{C. Niedermayer$^6$}

\author{Yusuke~Nomura$^8$}

\author{Ryotaro Arita$^{8,9}$}

\author{C. D. Batista$^{2,10}$}

\author{F. Ronning$^4$}

\author{M. Janoschek$^{1,4,7}$}

\affiliation{$^1$Laboratory for Neutron and Muon Instrumentation, Paul Scherrer Institute, Villigen PSI, Switzerland}%
\affiliation{$^2$Department of Physics and Astronomy, The University of Tennessee, Knoxville, TN, 37996, USA}%
\affiliation{$^3$School of Physics and Astronomy, University of Minnesota, Minneapolis, Minnesota 55455, USA}
\affiliation{$^4$Los Alamos National Laboratory, Los Alamos, NM 87545, USA}%
\affiliation{$^5$Neutron Scattering Division, Oak Ridge National Laboratory, Oak Ridge, Tennessee 37831, USA}
\affiliation{$^6$Laboratory for Neutron Scattering and Imaging, Paul Scherrer Institute, Villigen PSI, Switzerland}%
\affiliation{$^{7}$Physik-Institut, Universit\"{a}t Z\"{u}rich, Winterthurerstrasse 190, CH-8057 Z\"{u}rich, Switzerland}
\affiliation{$^8$RIKEN Center for Emergent Matter Science, Wako, Saitama 351-0198, Japan}
\affiliation{$^9$Department of Applied Physics, The University of Tokyo, Hongo, Bunkyo-ku, Tokyo 113-8656, Japan}
\affiliation{$^{10}$Quantum Condensed Matter Division and Shull-Wollan Center, Oak Ridge National Laboratory, Oak Ridge, TN, 37831, USA}
\affiliation{$^*$Present address: Center for Correlated Matter and School of Physics, Zhejiang University, Hangzhou 310058, China}

\maketitle

\clearpage

\section{Microscopic models}
The starting point for all materials specific models of $f$-electron systems is the multi-orbital periodic Anderson model (MO-PAM) $\mathcal{H}_{\text{MO-PAM}}$ presented in Eq.~(1) of the main text. To obtain the  parameters of $\mathcal{H}_{\text{MO-PAM}}$, we derive a tight-binding model based on density functional theory (DFT) that accurately captures the electronic structure of the conduction bands near the Fermi level $E_F$, and yields the hybridization between these bands and the $4f$-orbitals. We first performed the DFT band structure calculation for CeIn$_3$ using the \textsc{Quantum ESPRESSO} package~\cite{Gianozzi_2017} with the experimental lattice parameter: $a = 4.689$~\AA~\cite{Dijkman_1982}. We employed fully relativistic relativistic projector augmented-wave (PAW) pseudopotentials with the Perdew-Burke-Ernzerhof (PBE) exchange-correlation functional, which are available in the PSlibrary~\cite{Dal_Corso_2014}.
The calculation was performed with an 11$\times$11$\times$11 $\bf{k}$-mesh and an energy cutoff of 175 Ry for the wave function and 700 Ry for the charge density. 

\begin{figure*}
\includegraphics{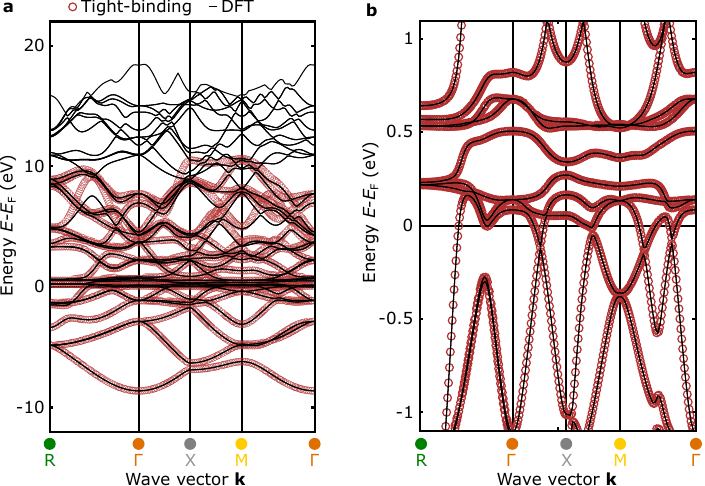}
\caption{\label{fig:DFT_TB_comparison_withf}
\textbf{Electronic band structure of CeIn$_3$.} Comparison of the DFT calculated band structure (solid lines) and the electronic structure of our 25-band tight-binding model (symbols). \textbf{a} Comparison including all the bands. \textbf{b} Comparision near $E_F$.}
\end{figure*}

To obtain a realistic tight-binding Hamiltonian, 50 Wannier functions (25 orbitals times 2 for spin) were constructed using the Wannier90 package~\cite{Pizzi_2020},
in which the projections of Ce $s$-, Ce $d$-, Ce $f$-, In $s$-, and In $p$-type orbitals were employed. 
We used an inner energy window of [$-1.5$ eV : +2 eV] with respect to the Fermi level ($E_{\rm F}$) to accurately reproduce the band structure around $E_{\rm F}$. The excellent agreement between our tight-binding model and the DFT computed band structure is shown in Fig.~\ref{fig:DFT_TB_comparison_withf}.

\begin{figure*}
\includegraphics{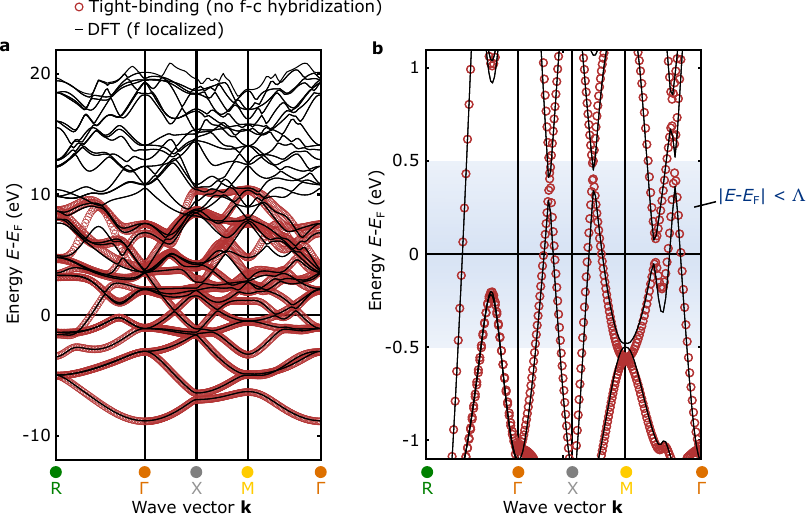}
\caption{\label{fig:DFT_TB_comparison_nof}
\textbf{Electronic band structure of CeIn$_3$ with localized $f$ electrons.} Comparison of the DFT calculated band structure with the $f$-electrons localized in the cores (solid lines) and the electronic structure of our 25 band tight-binding model with the hybridization between the $f$-orbitals (not shown) and the conduction bands (symbols) turned off. \textbf{a} Comparison including all the conduction bands.
\textbf{b} Comparison near $E_F$, where the shaded region corresponds to the cutoff $\Lambda=0.5$\,eV used in the calculation to separate the short- and long-range interactions.}
\end{figure*}

Notably, the DFT calculation and the resulting tight-binding model assume that the $f$-electrons are itinerant, whereas strong correlations will be responsible for localizing them. To illustrate that the bare conduction electron bands remain practically unchanged irrespective of the treatment of the $f$-electrons, in Fig.~\ref{fig:DFT_TB_comparison_nof} we present the DFT  band structure with the $f$-electrons localized in the core. We also include the bands obtained from the tight-binding model with itinerant $f$-electrons when setting the $f$-$c$ hybridization to zero $V_{\bf{k},\sigma s} = 0$. The good agreement illustrates that the bare conduction electron bands are insensitive to the treatment of the $f$-electrons. Furthermore, the quasi-flat nature of the resulting $f$-bands [bandwidth smaller than 0.03\,eV (0.07\,eV) for the $J=5/2$ ($J=7/2$) subbands] (shown in Fig.~2 of the main text) indicates that the $f$-$f$ hopping is negligible due to the small extension of the $4f$-orbitals. We note that the electronic band structure obtained in this way is also consistent with measurements of the electronic structure as demonstrated in section~\ref{sec:ARPES}.

From the first-principles calculations, we obtain the tight-binding model that includes 25 orbitals per spin. The low-temperature magnetic properties of CeIn$_3$ at zero field are dominated by the low-energy Ce 4$f$ $\Gamma_7$ doublet that results from diagonalizing the single-site $f$-electron Hamiltonian [second term of Eq.~(1) in the main text]:
\begin{subequations}
\begin{align}
|\Gamma_7;+\rangle &\equiv |\uparrow \rangle = \sqrt{\frac{1}{6}} \left|j=\frac{5}{2},m_j=\frac{5}{2}\right\rangle - \sqrt{\frac{5}{6}} \left|j=\frac{5}{2},m_j=-\frac{3}{2}\right\rangle,\\
|\Gamma_7;-\rangle &\equiv |\downarrow \rangle = \sqrt{\frac{1}{6}} \left|j=\frac{5}{2},m_j=-\frac{5}{2}\right\rangle - \sqrt{\frac{5}{6}} \left|j=\frac{5}{2},m_j=\frac{3}{2}\right\rangle.
\end{align}
\end{subequations}
By taking the limit of infinite intra-atomic $f$-$f$ Coulomb repulsion which eliminates the $f^2$ configurations, we project $\mathcal{H}_{\text{MO-PAM} }$ into the low-energy subspace generated by the $\Gamma_7$ doublet and obtain the following periodic Anderson model 
\begin{equation}
\mathcal{H}_{\rm PAM} = \sum_{\bf{k}, s} \epsilon_{\bf{k}, s} c_{\bf{k}, s}^{\dag} c_{\bf{k}, s}^{} + \epsilon_{\Gamma_7}^{f} \sum_{i,\sigma} \tilde f_{i, \sigma}^{\dag} \tilde f_{i,\sigma}^{} + \sum_{\bf{k}, \sigma,s} \left( \tilde{V}_{\bf{k},\sigma s} \tilde f_{\bf{k},\sigma}^{\dag} c_{\bf{k}, s}  + h.c. \right) ,
\label{PAM2}
\end{equation}
where the constrained operators $\tilde f_{i, \sigma}^{\dag}$ ($\tilde f_{i, \sigma}$) create (annihilate) an $f$-electron in the $\Gamma_7$ doublet with $\sigma=\{\uparrow,\downarrow \}$ and ${\tilde f}^{\dagger}_{i, \sigma} {\tilde f}^{\dagger}_{i, \sigma'}=0$, $\epsilon_{\Gamma_7}^{f}$ is the energy of the $\Gamma_7$ states, and $\tilde{V}_{\bf{k},\sigma s}$ is the hybridization between the $\Gamma_7$ doublet and the conduction electron states ($1 \leq s \leq 36$). $s$ is a combined spin and orbital index for the conduction electrons.

\begin{figure*}[tbp]
\includegraphics[]{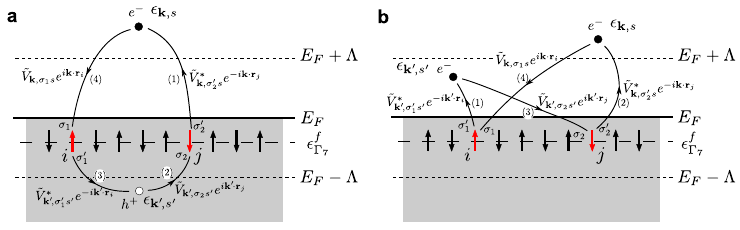}
\caption{\label{fig:ph-pp-diag}
Diagrams of the virtual processes $f^1 \rightarrow f^0 \rightarrow f^1$ in the particle-hole \textbf{a} and particle-particle \textbf{b} channels. Red and black short vertical arrows represent the spin ($\sigma$) of the localized electrons. The full (empty) dot represents a particle (hole) excitation with energy $\epsilon_{\bm k, s}$ ($\epsilon_{\bm k', s'}$). The numbers (1) to (4) indicate the sequence of virtual processes that contribute to the effective exchange interaction between $f$-moments to fourth order in the hybridization. Each process is indicated with a black line with an arrow in the middle that is accompanied by the amplitude of the process. The variables $\sigma_{1}$ and $\sigma_{2}$ ($\sigma_{1}^{'}$ and $\sigma_{2}^{'}$) represent the initial (final) states of the $f$-moments $i$ and $j$, respectively. The gray (white) region indicates the conduction band states below (above) the Fermi level.
Dashed lines indicate the energy cut-off that restricts the excited states that contribute to the RKKY interactions (ph within the cut-off $\Lambda$), the superexchange (ph and pp {involving at least one virtual state outside} the cut-off), and a long-range interaction $\hat K^{(\text{pp})}$ (pp within the cut-off).}
\end{figure*}

By treating the small hybridization $\tilde{V}_{\bf{k},\sigma s}$ as a perturbation, the periodic Anderson model can be reduced to an effective Kondo lattice model, which appears to second order in the hybridization, plus an effective exchange  interaction between the $\Gamma_7$ doublets that is generated to fourth order in the hybridization, which is necessary to properly account for the magnetic interactions as detailed below:
\begin{equation}
 \mathcal{H}_{\rm PAM} \rightarrow \mathcal{H}_{\rm KLM} + \mathcal{H}_{\rm Heis}.
 \label{KLMpHeis}
\end{equation}
The Kondo lattice model Hamiltonian is 
\begin{equation}
\mathcal{H}_\text{KLM}= \sideset{}{'}\sum_{\bf{k}, s}  \epsilon_{\bf{k},s} {c}_{\bf{k},s}^\dagger {c}_{\bf{k},s} + 
\sideset{}{'}\sum_{\substack{i,\bf{k}, \bf{k}^\prime,   \\
\sigma, \sigma^{\prime}, s, s^{\prime} }}
{J}_{i,\sigma \sigma^\prime}^{\bf{k}s,\bf{k}^\prime s^\prime} \tilde f_{i,\sigma}^\dagger \tilde f_{i, \sigma^\prime} {c}_{\bf{k}^\prime,s^\prime}^\dagger {c}_{\bf{k},s},
\label{KLModel}
\end{equation}
where the local Hilbert space is constrained by the condition $\sum_\sigma {\tilde f}_{i,\sigma}^\dagger {\tilde f}_{i,\sigma}=1$, which implies that the $f$-electrons  are localized in their orbitals and their effective spin $\sigma$ is the only remaining $f$-degree of freedom. It is important to note that we have only retained the conduction electron states whose distance to the Fermi level is smaller than a  cut-off $\Lambda$ for both the hopping and the Kondo coupling terms (indicated in the constrained sums). The Kondo interaction is then given by
\begin{equation}
{J}_{i,\sigma\sigma^{\prime}}^{\bf{k}s,\bf{k}^\prime s^{\prime}}
\approx
\frac{1}{N} \frac{\tilde{V}_{\bf{k},\sigma s} \tilde{V}_{\bf{k}^\prime,\sigma^{\prime}s^{\prime}}^{*} }{E_F-\epsilon_{\Gamma_7}^f} e^{\iu(\bf{k}-\bf{k}^\prime )\cdot\bf{r}_{i}} , \label{eq:Kondo_approx}
\end{equation}
where we have approximated the energy of the virtual processes $\epsilon_{\bf{k},s}-\epsilon_{\Gamma_7}^f$ by $E_F-\epsilon_{\Gamma_7}^f$.
The Heisenberg Hamiltonian ($\mathcal{H}_{\rm Heis}$) includes all magnetic interactions between the $\Gamma_7$ states generated through a fourth order degenerate perturbation theory of the periodic Anderson model Eq.~\eqref{PAM2}, which are not captured in the KLM of Eq.~\eqref{KLModel}. This includes all particle-particle (pp) processes at all energies (Fig.~\ref{fig:ph-pp-diag}), which are not included in the Kondo model \cite{1999_Fazekas}, and particle-hole (ph) processes that involve at least one virtual excited state outside the cut-off $|\epsilon_{\bf{k},s}-E_F| \geq \Lambda$.

For strongly localized $f$-electrons, the magnetic Ruderman-Kittel-Kasuya-Yosida (RKKY) interaction dominates the Kondo effect. The  Kondo lattice model can be further reduced to the RKKY  Hamiltonian via second order degenerate perturbation theory in the Kondo interaction (i.e. fourth order in the hybridization $\tilde{V}_{\bf{k},\sigma s}$). To properly account for all spin interactions from our MO-PAM that are up to fourth order in the hybridization we obtain a spin Hamiltonian as follows:
\begin{equation}
 \mathcal{H}_{\rm PAM} \rightarrow \mathcal{H}_{\rm spin} = \mathcal{H}_{\rm RKKY} + \mathcal{H}_{\rm Heis}
\end{equation}
where
\begin{equation}
\mathcal{H}_\text{RKKY}=\frac{1}{2}\sum_{i,j}\sum_{\sigma_{1},\sigma_{1}^{\prime},\sigma_{2},\sigma_{2}^{\prime}} {K}_{\sigma_{1}\sigma_{1}^{\prime}\sigma_{2}\sigma_{2}^{\prime}}^{\rm RKKY} (i,j)  \tilde f_{i,\sigma_{1}}^{\dagger}\tilde f_{i,\sigma_{1}^{\prime}}\tilde f_{j,\sigma_{2}}^{\dagger}\tilde f_{j,\sigma_{2}^{\prime}},
\end{equation}
and
\begin{equation}
 \mathcal{H}_{\rm Heis} = \frac{1}{2} \sum_{i,j} \sum_{ \sigma_1, \sigma_1^{\prime} , \sigma_2, \sigma_2^{\prime} } K_{\sigma_1 \sigma_1^{\prime}\sigma_2 \sigma_2^{\prime}}^{\rm Heis} (i,j) \ \tilde f_{i, \sigma_1}^{\dag} \tilde f_{i, \sigma_1^{\prime}} \tilde f_{j, \sigma_2}^{\dag} \tilde f_{j, \sigma_2^{\prime}} .
\end{equation}

$\mathcal{H}_{\rm spin}$ can also be derived directly from the periodic Anderson model using fourth order perturbation theory. In doing so, $\mathcal{H}_{\rm spin}$ includes all the ph and pp fourth order processes in the hybridization $\tilde{V}_{\bf{k},\sigma s}$, which are depicted in Fig.~\ref{fig:ph-pp-diag}. 
The contributions from the ph channel, shown in Fig.~\ref{fig:ph-pp-diag}\textbf{a}, produce  an effective interaction
\begin{align}
 K_{\sigma_1 \sigma_1'\sigma_2 \sigma_2'}^{(\text{ph})} (i,j) &= \sum_{ s , s' }  \int_{\cal{B}} \frac{\mathrm{d} \bf{k}}{(2\pi)^3} \int_{\cal{B}} \frac{\mathrm{d} \bf{k}^\prime}{(2\pi)^3}    \frac{ \tilde V_{{\bf{k}}, \sigma_2' s}^*  \tilde V_{{\bf{k}^\prime}, \sigma_2 s'}^{}  \tilde V_{{\bf{k}^\prime}, \sigma_1' s'}^{*} \tilde V_{{\bf{k}}, \sigma_1 s}^{} }{\epsilon_{\bf{k}^\prime,s'} - \epsilon_{\bf{k},s} }  \  e^{i ({\bf{k}} - {\bf{k}^\prime}) \cdot ({\bf{r}}_i - {\bf{r}}_j) }  \nonumber  \\
 &\quad \quad \quad \times \left[ \frac{ f(\epsilon_{\bf{k}^\prime,s'}) [1-f(\epsilon_{\bf{k},s})]  }{(\epsilon_{\bf{k},s} - \epsilon_{\Gamma_7}^f )^2 } - \frac{ f(\epsilon_{\bf{k},s}) [1-f(\epsilon_{\bf{k}^\prime,s'})]  }{(\epsilon_{\bf{k}^\prime,s'} - \epsilon_{\Gamma_7}^f )^2 }  \right] ,
\end{align}
where $\cal{B}$ stands for the first Brillouin zone.
While the contributions from the pp channel, shown in Fig.~\ref{fig:ph-pp-diag}\textbf{b},  produce an effective coupling
\begin{align}
 K_{\sigma_1 \sigma_1' \sigma_2 \sigma_2'}^{(\text{pp})} (i,j) &= \sum_{s , s' } \int_{\cal{B}} \frac{\mathrm{d} \bf{k}}{(2\pi)^3}  \int_{\cal{B}} \frac{\mathrm{d}  \bf{k}^\prime}{(2\pi)^3}    \tilde V_{{\bf{k}}, \sigma_1 s}^{}  \tilde V_{{\bf{k}^\prime}, \sigma_1' s'}^{*}  \tilde V_{{\bf{k}^\prime}, \sigma_2 s'}^{}   \tilde V_{{\bf{k}}, \sigma_2' s}^{*} \ e^{ i({\bf{k}} - {\bf{k}^\prime}) \cdot ({\bf{r}}_i - {\bf{r}}_j)} \nonumber \\
 & \quad \quad \quad\times \frac{ \epsilon_{\bf{k},s} + \epsilon_{\bf{k}^\prime,s'} - 2\epsilon_{\Gamma_7}^f  }{ \left(\epsilon_{\bf{k}^\prime,s'} - \epsilon_{\Gamma_7}^f \right)^2  \left(\epsilon_{\bf{k},s} - \epsilon_{\Gamma_7}^f \right)^2 }        \ [1-f(\epsilon_{\bf{k},s})] [1- f(\epsilon_{\bf{k}^\prime,s'})] .
\end{align}
The Fourier transform of these effective interactions is:
\begin{align}
 K_{\sigma_1 \sigma_1'\sigma_2 \sigma_2'}^{(\text{ph})} (\bf{q}) &=  \sum_{s, s'} \int_{\cal{B}} \frac{\mathrm{d} \bf{k}}{(2\pi)^3}    \frac{  \tilde V_{{\bf{q} + \bf{k}}, \sigma_1 s}^{} \tilde V_{{\bf{k}}, \sigma_1' s'}^{*}   \tilde V_{{\bf{k}}, \sigma_2 s'}^{} \tilde V_{{\bf{q} + \bf{k}}, \sigma_2' s}^* }{\epsilon_{\bf{k},s'} - \epsilon_{\bf{q} + \bf{k},s} }    \nonumber \\
 & \quad \quad \quad \times       \left[ \frac{ f(\epsilon_{\bf{k},s'}) [1-f(\epsilon_{\bf{q} + \bf{k},s})]  }{(\epsilon_{\bf{q} + \bf{k},s} - \epsilon_{\Gamma_7}^f )^2 } - \frac{ f(\epsilon_{\bf{q} + \bf{k},s}) [1-f(\epsilon_{\bf{k},s'})]  }{(\epsilon_{\bf{k},s'} - \epsilon_{\Gamma_7}^f )^2 }  \right],   \label{eq:Kph} \\
 K_{\sigma_1 \sigma_1'\sigma_2 \sigma_2'}^{(\text{pp})} (\bf{q}) &= \sum_{s, s'} \int_{\cal{B}}  \frac{\mathrm{d} \bf{k}}{(2\pi)^3}     \tilde V_{{\bf{q} + \bf{k}}, \sigma_1 s}^{}  \tilde V_{{\bf{k}}, \sigma_1' s'}^{*}  \tilde V_{{\bf{k}}, \sigma_2 s'}^{}   \tilde V_{{\bf{q} + \bf{k}}, \sigma_2' s}^{*}  \nonumber \\
 & \quad \quad \quad \times     \frac{ \epsilon_{\bf{q} + \bf{k},s} + \epsilon_{\bf{k},s'} - 2\epsilon_{\Gamma_7}^f  }{ \left(\epsilon_{\bf{k},s'} - \epsilon_{\Gamma_7}^f \right)^2  \left(\epsilon_{\bf{q} + \bf{k},s} - \epsilon_{\Gamma_7}^f \right)^2 }   \ [1-f(\epsilon_{\bf{q} + \bf{k},s})] [1- f(\epsilon_{\bf{k},s'})]. \label{eq:Kpp}
 \end{align}
In order to separate the contributions to $\mathcal{H}_{\rm RKKY}$ and $ \mathcal{H}_{\rm Heis}$, we must  split both interactions into two terms, $K^{(\text{ph})}(\bf{q}) = \hat K^{(\text{ph})}(\bf{q})+ \check K^{(\text{ph})}(\bf{q})$ [$K^{(\text{pp})}(\bf{q}) = \hat K^{(\text{pp})}(\bf{q})+ \check K^{(\text{pp})}(\bf{q})$], where for $ \hat K^{(\text{ph})}(\bf{q})$ [$ \hat K^{(\text{pp})}(\bf{q})$] the integration over momenta is restricted by the condition $|\epsilon_{\bf{k}, s'} - E_F|<\Lambda$ and $|\epsilon_{\bf{q} + \bf{k}, s} - E_F|<\Lambda$, with the cut-off $\Lambda = 0.5$ eV. The remaining contribution to the integral (at least one of the virtual states is outside the energy interval defined by the cut-off) is included in $ \check K^{(\text{ph})}(\bf{q})$ [$ \check K^{(\text{pp})}(\bf{q})$]. Now we can see that the RKKY contribution derived from the KLM [Eq.~\eqref{KLModel}] only includes processes that contribute to $ \hat K^{(\text{ph})}(\bf{q})$. The remaining three terms ($ \check K^{(\text{ph})}(\bf{q})$, $ \hat K^{(\text{pp})}(\bf{q})$, and $ \check K^{(\text{pp})}(\bf{q})$) are interactions not accounted for in our KLM. Consequently, the sum of these three terms are defined as our Heisenberg contribution, and must be added to our low energy effective model Eq.~\eqref{KLMpHeis} if we aim to accurately account for magnetic interactions.

To derive the single-magnon dispersion, it is convenient to express the effective  spin Hamiltonian in terms of the  the pseudo spin-$\frac{1}{2}$ operators $ \bf{S}_{i}\equiv\frac{1}{2}\tilde f_{i,\alpha}^{\dagger}\bf{\sigma}_{\alpha\beta}\tilde f_{i,\beta} $:
\begin{equation}
\mathcal{H}_\text{spin} = \frac{1}{2}\sum_{\bf{q}, \nu} \tilde I_{\bf{q}}^{\nu \nu} S_{\bf{q}}^\nu S_{-\bf{q}}^\nu + \frac{1}{2}\sum_{\bf{q}} \left(\tilde I_{\bf{q}}^{xy}S_{\bf{q}}^{x}S_{-\bf{q}}^{y}+\tilde I_{\bf{q}}^{yz}S_{\bf{q}}^{y}S_{-\bf{q}}^{z}+\tilde I_{\bf{q}}^{zx}S_{\bf{q}}^{z}S_{-\bf{q}}^{x}\right),
\end{equation}
where
\begin{subequations}
\begin{align}
\tilde I_{\bf{q}}^{xx} &= \left[K_{\uparrow\downarrow\downarrow\uparrow}(\bf{q})+K_{\downarrow\uparrow\uparrow\downarrow}(\bf{q})+K_{\uparrow\downarrow\uparrow\downarrow}(\bf{q})+K_{\downarrow\uparrow\downarrow\uparrow}(\bf{q})\right], \\
\tilde I_{\bf{q}}^{yy} &= \left[K_{\uparrow\downarrow\downarrow\uparrow}(\bf{q})+K_{\downarrow\uparrow\uparrow\downarrow}(\bf{q})-K_{\uparrow\downarrow\uparrow\downarrow}(\bf{q})-K_{\downarrow\uparrow\downarrow\uparrow}(\bf{q})\right], \\
\tilde I_{\bf{q}}^{zz} &= \left[K_{\uparrow\uparrow\uparrow\uparrow}(\bf{q})+K_{\downarrow\downarrow\downarrow\downarrow}(\bf{q})-K_{\uparrow\uparrow\downarrow\downarrow}(\bf{q})-K_{\downarrow\downarrow\uparrow\uparrow}(\bf{q})\right], \\
\tilde I_{\bf{q}}^{xy} &= 2\iu\left[-K_{\uparrow\downarrow\downarrow\uparrow}(\bf{q})+K_{\downarrow\uparrow\uparrow\downarrow}(\bf{q})+K_{\uparrow\downarrow\uparrow\downarrow}(\bf{q})-K_{\downarrow\uparrow\downarrow\uparrow}(\bf{q})\right], \\
\tilde I_{\bf{q}}^{yz} &=2\iu\left[K_{\uparrow\downarrow\uparrow\uparrow}(\bf{q})+K_{\downarrow\uparrow\downarrow\downarrow}(\bf{q})-K_{\uparrow\downarrow\downarrow\downarrow}(\bf{q})-K_{\downarrow\uparrow\uparrow\uparrow}(\bf{q})\right], \\
\tilde I_{\bf{q}}^{zx} &=2\left[K_{\uparrow\uparrow\uparrow\downarrow}(\bf{q})-K_{\downarrow\downarrow\downarrow\uparrow}(\bf{q})+K_{\uparrow\uparrow\downarrow\uparrow}(\bf{q})-K_{\downarrow\downarrow\uparrow\downarrow}(\bf{q})\right],
\end{align}
\end{subequations}
with $K(\bf{q}) = \hat K^{(\text{ph})}(\bf{q}) + \check K^{(\text{ph})}(\bf{q}) + \hat K^{(\text{pp})}(\bf{q}) + \check K^{(\text{pp})}(\bf{q}) $.

The effective spin-spin interaction turns out to be practically isotropic because of the absence of $f^2$ virtual states and the presence of the cubic symmetry:
\begin{equation}
\mathcal{H}_\text{spin} = \mathcal{H}_\text{RKKY} + \mathcal{H}_\text{Heis} \simeq \frac{1}{2}\sum_{\bf{q}, \nu} \tilde I_{\bf{q}}^{} S_{\bf{q}}^\nu S_{-\bf{q}}^\nu
\end{equation}
with $\tilde I_{\bf{q}}^{} = I_{\bf{q}}^{\rm RKKY} + I_{\bf{q}}$.
Indeed, our numerical evaluation confirms that $\tilde{I}_{\bf{q}}\equiv \tilde{I}_{\bf{q}}^{xx}=\tilde{I}_{\bf{q}}^{yy}=\tilde{I}_{\bf{q}}^{zz}$. Furthermore, the terms \{$\tilde{I}_{\bf{q}}^{xy}$, $\tilde{I}_{\bf{q}}^{yz}$, $\tilde{I}_{\bf{q}}^{zx}$\} that correspond to the cubic anisotropies are found to be much smaller than 0.01 meV.

\begin{figure*}[tbp]
\includegraphics{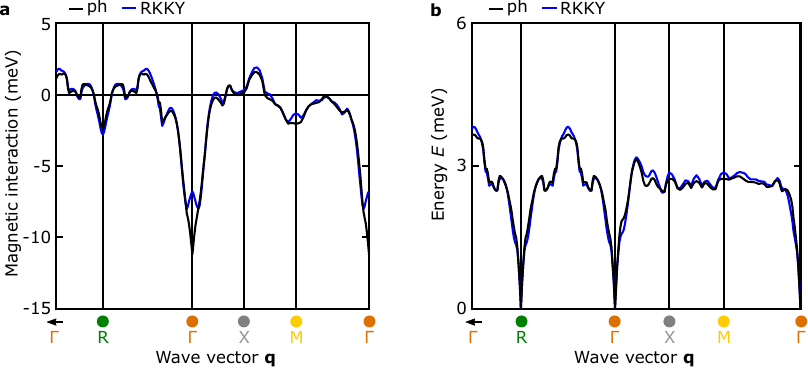}
\caption{\label{fig:S4} (a) Comparison between RKKY interactions $ \hat I_{\bf{q}}^{(\text{ph})} $ (black line) and $ I_{\bf{q}}^{\rm RKKY} $ (blue line) derived from the PAM and from the KLM, respectively. (b) 
Shown is the magnon dispersion on the path R$\Gamma$XM$\Gamma$ obtained directly from the PAM, i.e., using $ \hat I_{\bf{q}}^{(\text{ph})} $ (black line), as well as the dispersion resulting from the approximated RKKY interaction $ I_{\bf{q}}^{\rm RKKY} $ (blue line) that is obtained from the KLM. The same Heisenberg contribution is included for both magnon dispersions shown in (b).}
\end{figure*}

The RKKY Hamiltonian is
\begin{equation}
\mathcal{H}_\text{RKKY} = \frac{1}{2}\sum_{\bf{q}, \nu} I_{\bf{q}}^{\rm RKKY} S_{\bf{q}}^\nu S_{-\bf{q}}^\nu
\end{equation}
where $ I_{\bf{q}}^{\rm RKKY} $ has been derived from the Kondo lattice model. Due to the need to approximate the conduction electron particle hole states to lie at the Fermi energy, when deriving the Kondo lattice model, $ I_{\bf{q}}^{\rm RKKY} $ is similar, but not exactly equal, to $ \hat I_{\bf{q}}^{(\text{ph})} $. A comparison shown in Fig.~\ref{fig:S4}(a) illustrates the similarity, and hence, the validity in the approximation used to derive the Kondo lattice model. Fig.~\ref{fig:S4}(b) shows a comparison between the magnon dispersions obtained including either $ I_{\bf{q}}^{\rm RKKY} $ or $ \hat I_{\bf{q}}^{(\text{ph})} $ together with the Heisenberg contribution.

The Heisenberg term is
\begin{equation}
\mathcal{H}_\text{Heis} = \frac{1}{2}\sum_{\bf{q}, \nu} I_{\bf{q}} S_{\bf{q}}^\nu S_{-\bf{q}}^\nu
\end{equation}
with $I_{\bf{q}} \equiv I_{\bf{q}}^{\rm SE} + \hat I_{\bf{q}}^{(\text{pp})}$. The effective superexchange interaction $I_{\bf{q}}^{\rm SE} \equiv \check I_{\bf{q}}^{(\text{ph})} + \check I_{\bf{q}}^{(\text{pp})}$ involves at least one virtual excited state outside the energy interval defined by the cut-off $\Lambda$.

In the calculation, the position of the Fermi level, $E_F=12.588$ eV, is fixed by the number of non-$f$ electrons per unit cell. The only free parameter of  the theory is the diagonal energy $\epsilon_{\Gamma_{7}}^{f}$ of the $\Gamma_7$ doublet. By fitting the bandwidth of the measured single-magnon dispersion we obtain $\epsilon_{\Gamma_{7}}^{f}\approx 12.076$eV for the effective spin-spin interaction derived from the PAM (using $ \hat I_{\bf{q}}^{(\text{ph})} $) and $\epsilon_{\Gamma_{7}}^{f}\approx 12.009$ eV for the effective spin-spin interaction derived from the KLM (using $ I_{\bf{q}}^{\rm RKKY} $), which is the case presented in the main text.

\clearpage

\section{Experimental methods}
In the following, we elaborate on the experimental methods. After a description of the sample preparation, details of the neutron-spectroscopy experiments are presented.
\subsection{Sample preparation}
\label{Supplement:Section:SamplePreparation}
\begin{figure}[h!]
\includegraphics{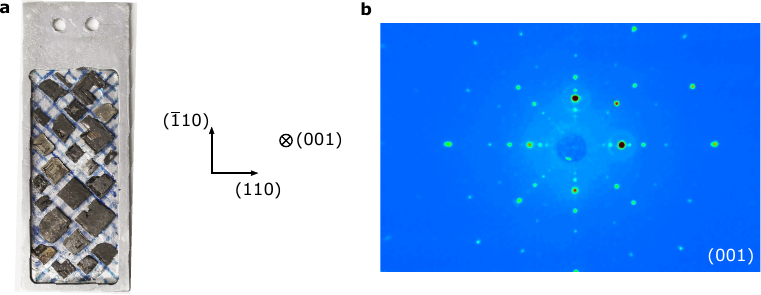}
\caption{\label{fig:S5}\textbf{a} Photograph of the CeIn$_3$ single-crystal mosaic. The crystallographic directions along that the crystals were coaligned are indicated on the right. \textbf{b} X-ray Laue diffraction pattern of one of the used CeIn$_3$ single-crystal samples illustrating the high quality. The data were recorded with the X-ray beam parallel to the $\left(001\right)$ crystallographic direction.}
\end{figure}

Figure~\ref{fig:S5} shows a photograph of the CeIn$_3$ sample-mosaic on the aluminum holder. The mosaic spread with respect to the $\left(\bar{1}10\right)$ rotation axis of the single-crystal pieces around the ideal orientation, which is indicated schematically on the right, is within $\pm1.4\,\mathrm{deg}$ rotation angle. Shown in \textbf{b} is an examplary Laue picture that indicates the high quality of one of the single-crystal pieces.

\subsection{Neutron spectroscopy experiments}
\label{Supplement:Section:NeutronSpectroscopy}

To determine the dispersion of magnetic excitations in CeIn$_3$ at $T=1.8\,\mathrm{K}$, inelastic neutron scattering (INS) experiments were performed using the time-of-flight technique at the cold neutron chopper spectrometer (CNCS) at Oak Ridge National Laboratory (ORNL)~\cite{2011_Ehlers_RevSciInstrum}. The orientation of the CeIn$_3$ sample mosaic for the experiment was such that the crystallographic direction $\left(1\bar{1}0\right)$ was vertical and $\left(110\right)$ as well as $\left(001\right)$ were in the horizontal scattering plane of CNCS (see Fig.~\ref{fig:S4}). 

The CNCS spectrometer was operated in the following two complementary instrumental settings:
\begin{itemize}
    \item In the first setting, which we refer to as high-energy setting, the incident neutron energy was $E_{i,1}=12\,\mathrm{meV}$ (wave vector $K_{i,1}=2.406\,\text{\AA}^{-1}$) and the choppers of CNCS were operated in the high-flux mode with $180\,\mathrm{Hz}$ double-disk rotation frequency. 
    \item In the second setting, which we refer to as high-resolution setting, the incident neutron energy was $E_{i,2}=3.315\,\mathrm{meV}$ (wave vector $K_{i,2}=1.265\,\text{\AA}^{-1}$) and the choppers were operated in the high-flux mode with $300\,\mathrm{Hz}$ double-disk rotation frequency.
\end{itemize}

The first setting provides an overview of the magnetic excitations in a large area of reciprocal space and up to high energy transfers, whereas the second setting permits high resolution measurements of the magnon dispersion in the vicinity of the momentum transfer $\bf{Q}_0=\left(-\frac{1}{2},-\frac{1}{2},\frac{1}{2}\right)$, which corresponds to probing excitations at the magnetic propagation vector $\bf{q}_\text{AFM}=\left(\frac{1}{2},\frac{1}{2},\frac{1}{2}\right)$~\cite{1980_Lawrence_PhysRevB} located at the R point.

For the data sets in the high-energy and high-resolution setting, respectively, the angular coverage of sample rotation amounted to 33 deg and to 50 deg with a step size of 1 deg.

\begin{figure}[h!]
\includegraphics{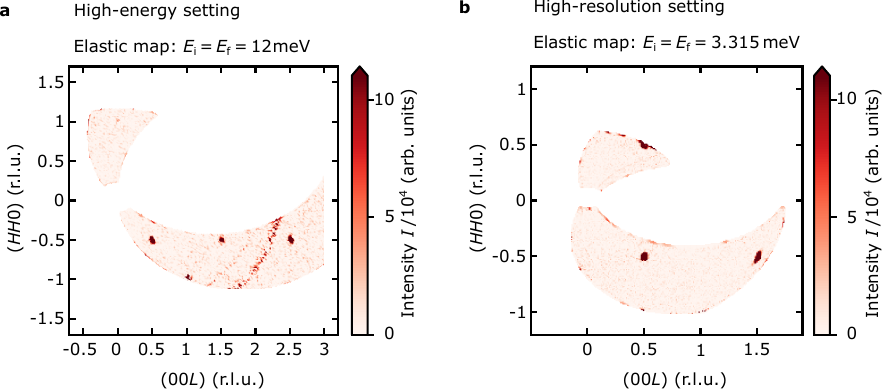}
\caption{\label{fig:S6}
\textbf{Maps of elastic neutron scattering intensity recorded in the horizontal scattering plane.} 
{\bf a} In the high-energy setting a large area of reciprocal space is accessible. Clearly visible are magnetic Bragg peaks at the momentum transfers, $\bf{Q}=$ $\left(-0.5,-0.5,0.5\right)$, $\left(-0.5,-0.5,1.5\right)$ and $\left(-0.5,-0.5,2.5\right)$ as well as a relatively weak temperature dependent signal at the structural positions $\left(1,1,0\right)$ and $\left(-1,-1,1\right)$. {\bf b} In the high-resolution setting the accessible range in reciprocal space is restricted to the close vicinity around the magnetic Bragg peak position $\left(-0.5,-0.5,0.5\right)$. Both panels show background-subtracted neutron scattering intensity that was integrated over the energy range $\Delta E=\pm 0.1 \,\mathrm{meV}$.}
\end{figure}

Figure~\ref{fig:S6} illustrates the areas of reciprocal space that were covered in the high-energy setting (cf. \textbf{a}) and in the high-resolution setting (cf. \textbf{b}) showing maps of elastic neutron-scattering intensity in the horizontal scattering plane. 

The pronounced Gaussian profile of the magnetic Bragg peak at $\left(-0.5,-0.5,0.5\right)$, as recorded in the high-resolution setting, indicates an excellent sample mosaicity with respect to the $\left(\bar{1}10\right)$ axis.

INS-data were recorded at $1.8\,\mathrm{K}$ (foreground) and at $20\,\mathrm{K}$ (background), i.e., in the magnetically ordered state and well above the ordering temperature. At each temperature, a horace scan was performed, i.e., the sample was rotated around the vertical axis. Data reduction was performed with the software \textit{HORACE}~\cite{2016_Ewings_NuclearInstrumentsandMethodsinPhysicsResearchSectionAAcceleratorsSpectrometersDetectorsandAssociatedEquipment}. The size of the background-corrected  magnetic intensity, $I(\bf{Q},\omega)$, reflects the dynamic magnetic scattering cross section, which is given by:
\begin{equation}
\mathrm{d}^2\sigma/\mathrm{d}\Omega\mathrm{d}\omega= \frac{N}{\hbar} \frac{K_f}{K_i}\left(\frac{g\gamma r_0}{2}\right)^2 \exp{(-2W)} F^2(\bf{Q})\sum_{\alpha,\beta}\left(\delta_{\alpha\beta}-\hat{Q}_{\alpha}\hat{Q}_{\beta}\right)S^{\alpha\beta}(\bf{Q},\omega) \, ,
\end{equation}
where $S^{\alpha\beta}$ denotes thy dynamic magnetic spin-correlation function, $\exp(-2W)$ denotes the Debye-Waller factor, $F$ the magnetic form factor, $N$ the number of unit cells, and $\frac{\gamma r_0}{2}$ the magnetic scattering length.

The dynamic magnetic correlation function is directly related to the imaginary part of the dynamic magnetic susceptibility, $\chi^{\alpha\beta}(\bf{Q},\omega)$, by the fluctuation-dissipation theorem:
\begin{align}
\chi^{\alpha\beta}{}''(\bf{Q},\omega) = g^2 \mu_{\mathrm{B}}^2 \frac{\pi}{\hbar}\left(1-\exp{\left(-\frac{\hbar\omega}{k_{\mathrm{B}}T}\right)}\right) S^{\alpha\beta}(\bf{Q},\omega)
\end{align}

Detailed accounts one these relations are found, e.g., in Refs.~\cite{2015_Janoschek_SciAdva,2013_Xu_RevSciInstrum}. An introduction into the theory of magnetic neutron scattering is further provided by Ref.~\cite{1984_Lovesey_Vol2}.

Setting $\tilde{\chi}{}''(\bf{Q},\omega):=\sum_{\alpha,\beta}\left(\delta_{\alpha\beta}-\hat{Q}_{\alpha}\hat{Q}_{\beta}\right)\chi^{\alpha\beta}{}''(\bf{Q},\omega)$ we find further~\cite{1984_Lovesey_Vol2,2015_Janoschek_SciAdva,2013_Xu_RevSciInstrum}:

\begin{equation}
\tilde{S}(\bf{Q},\omega)=\frac{13.77b^{-1}\tilde{I}(\bf{Q},\omega)}{g^2 F^2\exp{(-2W)}Nk_fR_0} \, 
\end{equation}
where $\tilde{I}(\bf{Q},\omega)$ denotes the time-of-flight scattering intensity.

To present scattering intensity on an absolute scale, we followed the procedure explained in Ref.~\cite{2013_Xu_RevSciInstrum}, normalizing the spectroscopy data to the incoherent scattering of the sample, $I_{ic}=\int \tilde{I}(\bf{Q},\omega)\mathrm{d}E$ using the relation:
\begin{equation}
Nk_fR_0=4\pi \frac{I_{ic}}{\sum_j \sigma^{ic}e^{-2W_j}}       \, .
\end{equation}

The experimentally observed incoherent scattering contains on the one hand contributions due to the sample, which are dominated by the scattering off indium having the cross section $\sigma^{ic}=0.54\cdot b$~\cite{1992_Sears_NeutronNews}. On the other hand, incoherent scattering arises from the aluminium sample holder. We assumed that the entire CeIn$_3$ sample and all aluminum in the sample holder contributed to the observed incoherent scattering. To assess the uncertainty in this normalization procedure, we assumed that we may have overestimated the amount of aluminum in the beam by 70\,\%. In addition, we considered that the procedure introduced in Ref.~\cite{2013_Xu_RevSciInstrum} is generally expected to suffer uncertainties of the order of 20\,\% on the resulting values due to systematic errors. 

The resolution volume $R_0$ as a function of energy was calculated by the \textit{Violini}-method using the software \textit{Takin} (see Refs.~\cite{2014_Violini_NuclearInstrumentsandMethodsinPhysicsResearchSectionAAcceleratorsSpectrometersDetectorsandAssociatedEquipment,2016_Weber_SoftwareX}) for the high-energy setting in the entire energy-range and for the high-resolution setting below 1.4 meV energy transfers. For the high-resolution measurements above 1.4\,meV energy transfer we assumed a constant resolution, as the \textit{Violini}-method failed.

\begin{figure}[h!]
\includegraphics{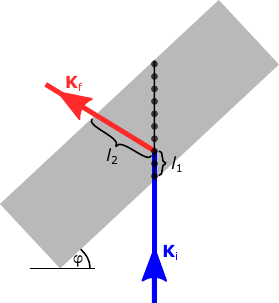}
\caption{\label{fig:S7}\textbf{Schematic illustration of the absorption correction.} The single-crystal mosaic comprises of plate-like samples for that the in-plane dimensions are much larger than their thickness. For a given incident wave-vector $\bf{K}_i$ (blue arrow), each momentum and energy transfer corresponds to a unique combination of scattered wave-vector $\bf{K}_f$ (red arrow) and rotation angle $\phi$ of the sample. For each such combination, the path of unscattered neutrons through the sample (gray rectangle), as indicated by a black thin line, was divided into 10 pieces (black circles) that scatter neutrons with equal probability. We determined the average absorption for the neutron paths $l_1$ and $l_2$ within the sample.}
\end{figure}

The neutron spectroscopy data were further corrected for absorption effects. Using a ray-tracing technique, a comprehensive sample-geometry dependent absorption correction due to indium, which in our case dominates the absorption, was performed. The required code for the absorption correction is available at \url{https://github.com/wolfgangsimeth/NeutronAbsorptionPlatelikeSamples}. Here the angle and energy dependent path of neutrons passing through the sample was considered in order to determine the intensity reduction due to  the absorption. For this the samples shown in Fig.~\ref{fig:S5} were estimated to be perfectly plate-like with an average thickness 0.78 mm. In the entire relevant parameter range, incident neutrons hit the platelets at one of their large facets, scatter, and exit the platelet on the opposite site (see Fig.~\ref{fig:S7}). The absorption cross sections on the paths $l_1$ and $l_2$ depend on the wave-lengths before and after scattering. The tabulated value for indium in Ref.~\cite{1992_Sears_NeutronNews},  ${\sigma}^{a}=193.8\cdot b$, refers to neutrons of wave-length $1.798\,\AA$. We assumed a linear dependence of ${\sigma}^{a}$ on the incident neutron wavelength $\lambda$. As a further constraint we limited the absorption coefficient to 0.1 in order to avoid divergences arising from specific constellations, where the scattered beam is parallel to the sample surface. These points corresponded to momentum and energy transfers that are far away from the settings in that the magnon dispersions in Fig.~3 and Fig.~4 of the main text was recorded.

To compare experiments with the dynamic spin correlation function $S^{\alpha.\beta}(\boldsymbol{Q},\omega)$ from MO-PAM calculations, we calculated the dynamic magnetic susceptibility for $g=\frac{6}{7}$. A comparison between experiment and theory on absolute scale is illustrated in Fig.~3 of the main text. The agreement is excellent, and even reproduces fine details in the structure of the intensity distribution (see. Fig.~3 (c) and (d) in the main text), in particular, when the size of the systematic errors in the conversion to absolute units (see above) as well as uncertainties in the calculation of the dynamic susceptibility are taken into account. Notably, for the calculation of the dynamic magnetic susceptibility, we note that additional effects such as frustration, Kondo screening, or quantum fluctuations can reduce the magnetic moment and the resulting dynamic magnetic susceptibility substantially. In the related material CeRhIn$_5$, we found that these effects can be about 20~\% \cite{2017_Fobes_JCPM}.

\clearpage

\section{Magnon dispersion inferred from experimental data}

In the following, we present a detailed account of the magnetic excitations in CeIn$_3$ as derived from our INS experiments. At first, we show neutron spectroscopy data that are complementary to the data in the main text. Subsequently, the excitations at the R point are studied in detail and the presence of a putative gap, previously reported by Knafo \textit{et al.} \cite{2003_Knafo_JPhysCondensMatter}, is discussed. Finally, the trajectory of the magnetic dispersion throughout the Brillouin zone is inferred from our data.

\subsection{Complementary INS data}
\begin{figure}[h!]
\includegraphics{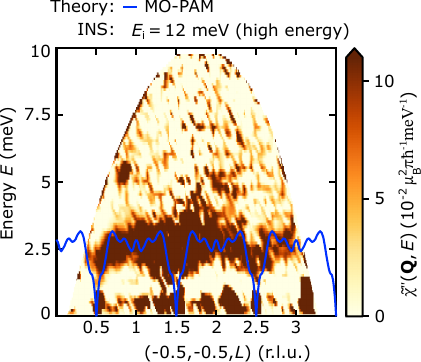}
\caption{\label{fig:S8}\textbf{Magnon dispersion on the path RM.} The imaginary part of the dynamic susceptibility, $\tilde{\chi}''(\bf{Q}, E)$, inferred from our experiments (see text for details) is illustrated. Intensity was integrated over a $Q$-range of distance $0.173\,\mathrm{r.l.u.}$ along the direction $\left(1\bar{1}0\right)$. The blue solid line denotes the magnon dispersion calculated via the MO-PAM.}
\end{figure}

Figure~\ref{fig:S8} presents neutron spectroscopy data of magnetic excitations on the path RM and over several Brillouin zones.

\subsection{Magnetic excitations at the R point}

The INS data recorded in the high-resolution setting reveal a relatively steep slope of the magnon dispersion emerging from the magnetic Bragg peak at $\bf{q}_{\mathrm{AFM}}=\left(\frac{1}{2},\frac{1}{2},\frac{1}{2}\right)$, as presented in Fig.~4 of the main text. As elaborated on in the following, the data indicate further, that a gap at the R point is either absent or smaller than the energy resolution of CNCS. In particular, the data pose an upper bound on the size of a putative gap at the R point. 

The instrumental energy resolution may be inferred from the incoherent scattering intensity at a $\bf{Q}$-position where elastic magnetic Bragg scattering is absent. The energy-dependent foreground intensity at $\bf{Q}_1=\left(-\frac{1}{2},-\frac{1}{2},\frac{1}{2}-0.1\right)$, which is obtained by integration of time-of-flight intensity over a cuboid $\bf{Q}$-volume centered at $\bf{Q}_1$ and extended over a distance of $\pm0.018\,\mathrm{r.l.u.}$ along the directions $\left(110\right)$, $\left(1\bar{1}0\right)$, and $\left(001\right)$ displays a Gaussian profile around the elastic line given by:
\begin{align}
    G\left(E\right)= I_0 \cdot \exp\left(-\frac{\left(E-E_0\right)^2}{2\sigma^2}\right)\,.
\end{align}
The width of the profile, which is characterised by $\sigma=45\,\mathrm{\mu e V}$ and a full width at half maximum $\mathrm{FWHM}=2\sqrt{2\ln2}\sigma=106\,\mathrm{\mu e V}$, essentially represents the resolution of CNCS in the high-resolution setting at zero energy transfer. The value is close to resolution of CNCS reported in literature (cf. Ref.~\cite{2011_Ehlers_RevSciInstrum,2017_Anastasopoulos_JInst}) and considerably smaller than $196\,\mathrm{\mu e V}$, which is the FWHM for incoherent Vanadium scattering estimated by the \textit{Violini}-algorithm by means of the software \textit{Takin} (see Ref.~\cite{2016_Weber_SoftwareX}). 

\begin{figure}[h!]
\includegraphics{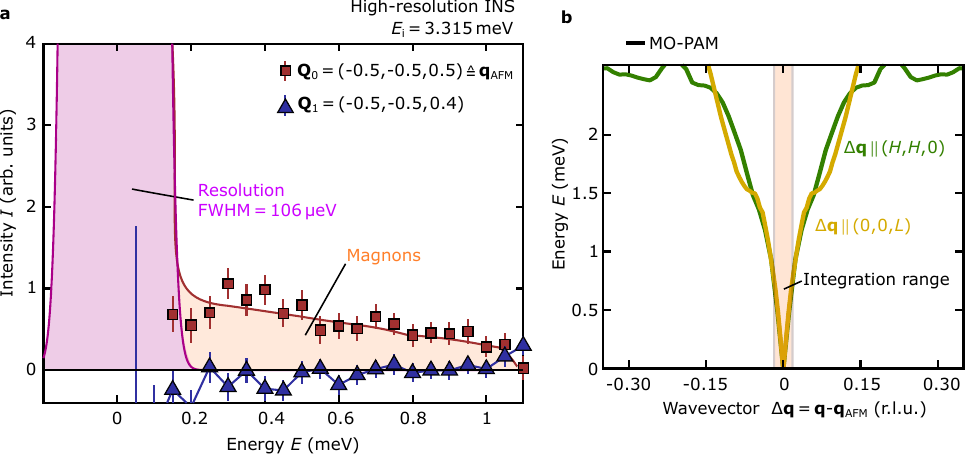}
\caption{\label{fig:S9}\textbf{Gapless magnon dispersion with extremely high slope at the $R$ point.} \textbf{a} INS intensity at the reciprocal space position $\bf{Q}_0$ (brown square symbols connected by a guide for the eyes), which corresponds to the R point, indicates finite intersection with the magnon dispersion up to energies around $1\,\mathrm{meV}$. The resulting neutron intensity is indicated by orange shading. In contrast, the scattering intensity at $\bf{Q}_1$ (blue triangle symbols connected by lines), which is slightly shifted from the $R$ point, is vanishingly small in the entire energy-transfer range. The plotted intensity was obtained by integrating over a $\bf{Q}$-volume extended over a distance of $\pm0.018\,\mathrm{r.l.u.}$ along the directions $\left(110\right)$, $\left(1\bar{1}0\right)$, and $\left(001\right)$. Error bars denote the statistical error. The Gaussian profile with pink shading corresponds to the energy resolution of the instrument. \textbf{b} Schematic view of the experimental integration range at $\bf{Q}_0$ compared to the magnon dispersion as inferred from MO-PAM calculations. The solid lines correspond to the calculated magnon dispersion along the cubic face diagonal (green) and the cubic edge (yellow). The integration range around $\bf{Q}_0$ is indicated by the brown shading.}\end{figure}

The R point was investigated at the momentum transfer $\bf{Q}_0=\left(-\frac{1}{2},-\frac{1}{2},\frac{1}{2}\right)$. Figure~\ref{fig:S9}\textbf{a} presents background-subtracted INS data as a function of energy transfer $E$ at the position $\bf{Q}_0$, where the elastic magnetic Bragg peak is located, as well as at the shifted position $\bf{Q}_1$, where magnetic intensity is absent. The magnetic Bragg peak at $\bf{Q}_0$ as a function of energy, $E$, displays a Gaussian profile with the same width as the instrumental resolution. The magnon dispersion intersects due to its steep slope with the integration dome of $\bf{Q}_0$ up to energies around $1\,\mathrm{meV}$ (see Fig.~\ref{fig:S9}\textbf{b}).

The onset of substantial scattering intensity at energy transfers right above the elastic Bragg peak indicates that the dispersion is gapless within the energy resolution of CNCS.

\subsection{Trajectory of Magnon Dispersion}

The trajectory in reciprocal space of the magnon dispersion was inferred from INS data from the high-energy and high-resolution setting, as explained in the following. 

Sufficiently far away from the points R and $\Gamma$ and in major parts of the Brillouin zone, the magnetic dispersion is relatively flat and the trajectory is clearly visible in constant $\bf{Q}$-cuts through the INS data that were recorded in the high-energy setting.

\begin{figure}[h]
\includegraphics[width=0.95\columnwidth]{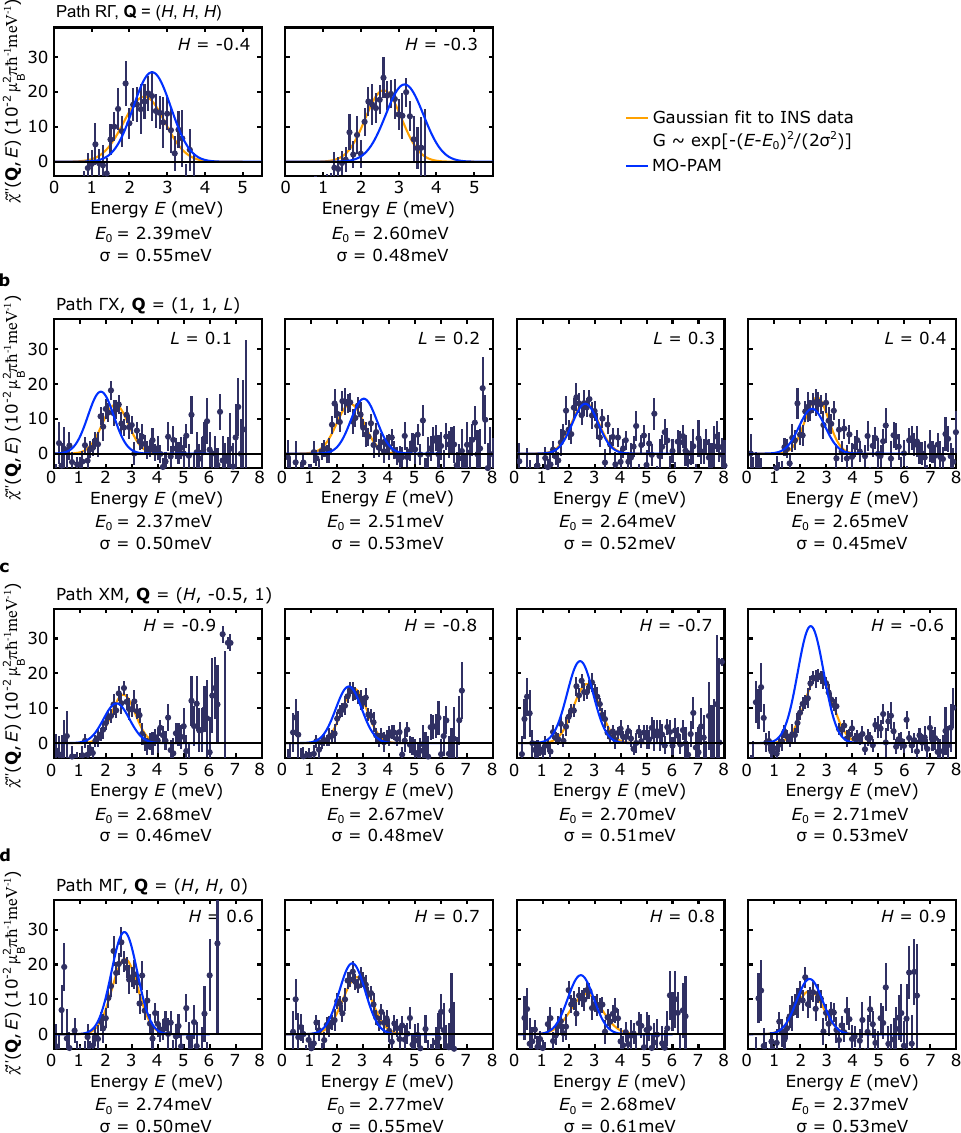}
\caption{\label{fig:S10}\textbf{Energy cuts through $\tilde{\chi}''(\bf{Q}, E)$, as inferred from the INS data set recorded in the high-energy setting, at constant $\bf{Q}$-positions distributed over the entire Brillouin zone.} \textbf{a}, \textbf{b}, \textbf{c}, and \textbf{d} present  cuts at $\bf{Q}$-positions that lie on the paths R$\Gamma$, $\Gamma$X, XM, and M$\Gamma$, respectively. The intersections with the magnon dispersion are characterised by Gaussian profiles (center $E_0$, FWHM $2\sqrt{2\ln2}\sigma$). Each data point on the path $\Gamma XM\Gamma$ represents intensity that was integrated over a distance of $\pm0.17\,\mathrm{r.l.u.}$ along three perpendicular $\bf{Q}$-directions. Error bars denote the statistical error. On the path $R\Gamma$ the integration ranges were $\pm0.09\,\mathrm{r.l.u.}$ and $\pm0.17\,\mathrm{r.l.u.}$ for $(\bar{1}\bar{1}1)$ and the other axes, respectively. On M$\Gamma$ the integration range was $\pm0.10\,\mathrm{r.l.u.}$ and $\pm0.17\,\mathrm{r.l.u.}$ for $(\bar{1}\bar{1}0)$ and the perpendicular axes, respectively. The solid orange and blue lines represent a Gaussian fitted to the data and the intensity distribution as calculated by the MO-PAM, respectively.}
\end{figure}

Figure~\ref{fig:S10} presents high-energy INS-intensity as a function of energy at several $\bf{Q}$-positions on the paths R$\Gamma$, $\Gamma$X, XM, and M$\Gamma$. The peaks that are visible indicate the intercepts with the magnon dispersion and were fitted with Gaussian profiles. Their centers indicate the location of the dispersion trajectory.

\begin{figure}[h]
\includegraphics{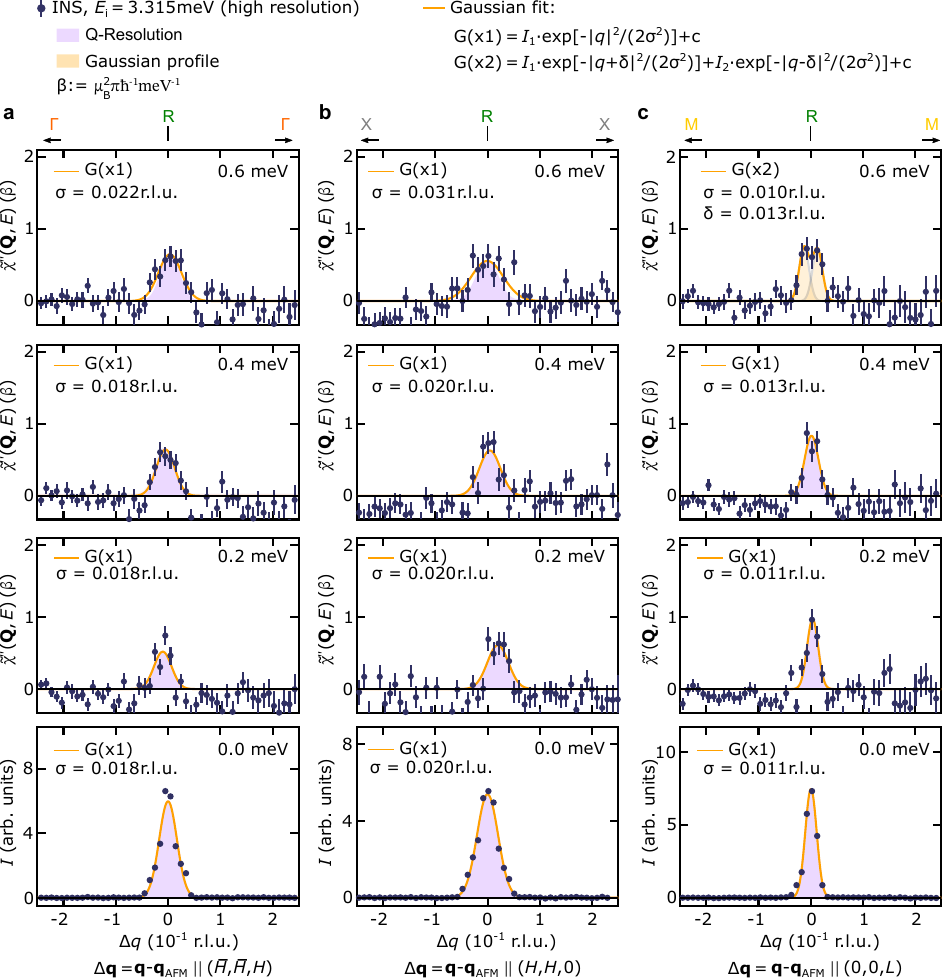}
\caption{\label{fig:S11}
\textbf{Magnetic excitations in the vicinity of the antiferromagnetic ordering vector at fixed energies ($0\,\mathrm{meV}\leq E\leq 0.6\,\mathrm{meV}$).} Shown are cuts through high-resolution INS-data at constant energies as a function of $\Delta q$, whereby $\bf{q}=\bf{q}-\bf{q}_{\mathrm{AFM}}$ is directed along the \textbf{a} cubic space diagonal R$\Gamma$, \textbf{b} cubic face diagonal RX, and \textbf{c} cubic edge RM, respectively. Error bars denote the statistical error. The data points that are shown were recorded at momentum transfers around $\bf{Q}=(-0.5,-0.5,0.5)$.}
\end{figure}

\begin{figure}
\includegraphics{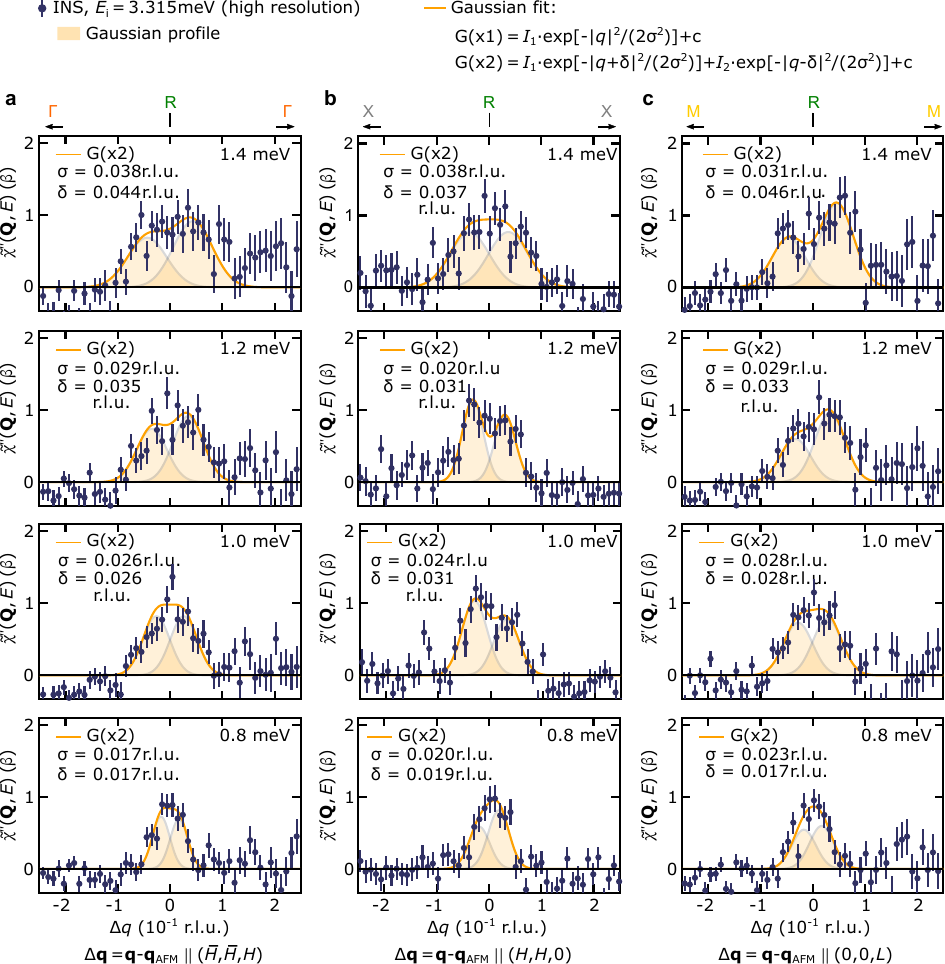}
\caption{\label{fig:S12}
\textbf{Magnetic excitations in the vicinity of the antiferromagnetic ordering vector at fixed energies ($0.8\,\mathrm{meV}\leq E\leq 1.4\,\mathrm{meV}$).} Shown are cuts through high-resolution INS-data at constant energies as a function of $\Delta q$, whereby $\bf{q}=\bf{q}-\bf{q}_{\mathrm{AFM}}$ is directed along the \textbf{a} cubic space diagonal R$\Gamma$, \textbf{b} cubic face diagonal RX, and \textbf{c} cubic edge RM, respectively. Error bars denote the statistical error. The data points that are shown were recorded at momentum transfers around $\bf{Q}=(-0.5,-0.5,0.5)$.}
\end{figure}

\begin{figure*}
\includegraphics{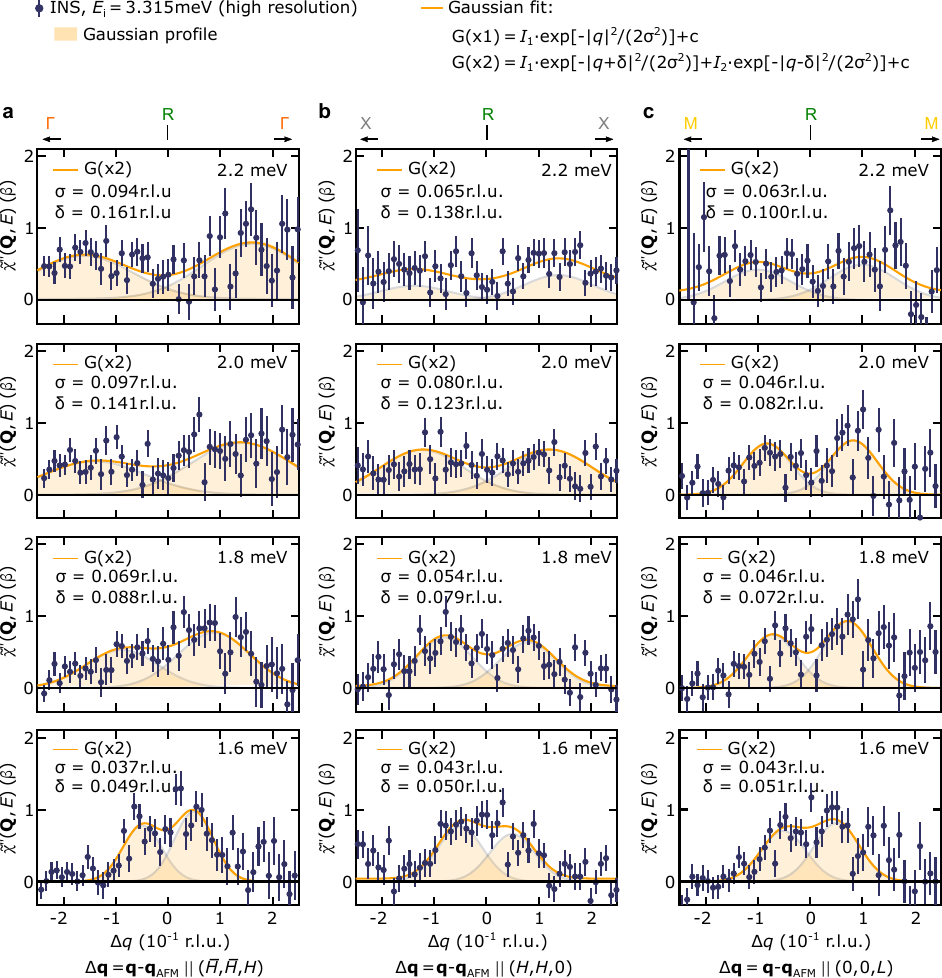}
\caption{\label{fig:S13}
\textbf{Magnetic excitations in the vicinity of the antiferromagnetic ordering vector at fixed energies ($1.6\,\mathrm{meV}\leq E\leq 2.2\,\mathrm{meV}$).} Shown are cuts through high-resolution INS-data at constant energies as a function of $\Delta q$, whereby $\bf{q}=\bf{q}-\bf{q}_{\mathrm{AFM}}$ is directed along the \textbf{a} cubic space diagonal R$\Gamma$, \textbf{b} cubic face diagonal RX, and \textbf{c} cubic edge RM, respectively. Error bars denote the statistical error. The data points that are shown were recorded at momentum transfers around $\bf{Q}=(-0.5,-0.5,0.5)$..}
\end{figure*}

Close to the R point, where the magnetic ordering vector is located, the gapless magnetic excitation dispersion displays a steep slope and the trajectory is visible in cuts at constant energy transfers through the high-resolution INS data set. Figs.~\ref{fig:S11}, \ref{fig:S12}, and \ref{fig:S13} present constant-energy cuts for momentum transfers around $\bf{Q}_0=\left(-\frac{1}{2},-\frac{1}{2},\frac{1}{2}\right)$. For the cuts at finite energies, i.e., $E>0$, the data were corrected by the Bose-factor and the resulting intensity represents the imaginary part of the susceptibility at constant E.

As a function of $\Delta\bf{q}=\bf{q}-\bf{q}_{\mathrm{AFM}}$ along the three reciprocal-space directions $\left(\bar{H}\bar{H}H\right)$, $\left(\bar{H}\bar{H}0\right)$, and $\left(00L\right)$ the magnon dispersion features two intersections in constant-energy cuts taken at energies $\leq 2.2\,\mathrm{meV}$. Due to the reciprocity of spin waves the intersections are symmetric around the center $\bf{q}_{\mathrm{AFM}}$, i.e., they appear at values $+\bf{q}$ and $-\bf{q}$, but possibly with different spectral weights. 

At the lowest energies, the separation of intersection points is smaller than the FWHM of the experimental momentum-transfer-resolution (purple shading) and the cuts are well fitted by a single-Gaussian profile (orange shading) denoted $1G$. For higher energies the splitting exceeds the FWHM of the instrumental resolution, i.e., $2\delta>f_{\mathrm{res}}$, and the profile can better be fitted by two Gaussian peaks being symmetric around the center.

The width as a function of $\bf{Q}$ of the magnetic Bragg peak at momentum $\bf{Q}_0=\left(-\frac{1}{2},-\frac{1}{2},\frac{1}{2}\right)$ (zero energy transfer)
essentially represents the instrumental resolution convoluted with the mosaicity of the invesitgated sample, which for the directions $\left(\bar{H}\bar{H}H\right)$, $\left(\bar{H}\bar{H}0\right)$, and $\left(00L\right)$ is described by a Gaussian profile with FWHM of $f_{(\bar{H}\bar{H}H)}\approx 0.054\,\text{\AA}^{-1}$, $f_{(HH0)}\approx 0.063\,\text{\AA}^{-1}$, and $f_{(00L)}\approx 0.035\,\text{\AA}^{-1}$, respectively. The values indicate an excellent mosaicity of the sample array that is smaller than $\pm1.4\,\mathrm{deg}$ around the vertical axis. The ratio for the instrumental resolution along $(\bar{H}\bar{H}H)$ and $(00L)$ is in agreement with the values inferred with the \textit{Violini}-algorithm by means of the \textit{Takin} software (see Ref.~\cite{2016_Weber_SoftwareX}), where $f^{\mathrm{Violini}}_{(\bar{H}\bar{H}H)}\approx 0.027\,\text{\AA}^{-1}$ and $f^{\mathrm{Violini}}_{(00L)}\approx 0.015\,\text{\AA}^{-1}$ were obtained. 

Experimental data points of magnon trajectories that are presented in Figs.~3 and 4 of the manuscript were inferred from the Gaussian fit profiles that were presented in this section and the errorbars of data points represent one $\sigma$ of the fitted Gaussian profiles.

\clearpage

\section{Magnon velocity}
The magnon group velocity at ordering vector $\bf{q}_{\mathrm{AFM}}=\left(\frac{1}{2},\frac{1}{2},\frac{1}{2}\right)$ along a direction $\Delta\bf{q}=\bf{q}-\bf{q}_{\mathrm{AFM}}$ is given by:
\begin{equation}
v_{\Delta\bf{q}} = \hat{e}_{\Delta\bf{q}} \cdot \left. \nabla E_{\bf{k}} \right|_{\bf{k}=\bf{q}_{\mathrm{AFM}}},
\end{equation}
where $\hat{e}_{\Delta\bf{q}}\equiv \Delta\bf{q}/|\Delta\bf{q}|$.

Magnon velocities at the R point, i.e., $\bf{q}_{\mathrm{AFM}}=\left(\frac{1}{2},\frac{1}{2},\frac{1}{2}\right)$, were determined by linear fits to high-resolution INS data with momentum around $\bf{Q}_{0}=\left(-\frac{1}{2},-\frac{1}{2},\frac{1}{2}\right)$ and compared with computations.

\subsection{Magnon velocities inferred from experimental data}

Magnon velocities were determined for the three cases, where $\Delta\bf{q}$ is directed along the cubic space diagonal R$\Gamma$, the cubic face diagonal RX, and the cubic edge RM. The dispersions along the three directions are presented in Fig.~4 of the manuscript. At small wave vectors $\Delta\bf{q}$ with respect to the R point and energy transfers $\leq 1.6\,\mathrm{meV}$ the magnetic excitations display essentially a linear dispersion. 

At lowest energies, where the distance of the two intersection points in constant-energy cuts as a function of $\Delta q=|\bf{q}-\bf{q}_{\mathrm{AFM}}|$ (see Figs.~\ref{fig:S11}, \ref{fig:S12}, and \ref{fig:S13}) is smaller than the FWHM of the instrumental resolution, an estimate for the magnon velocity is obtained from the resolution. For the cubic space diagonal R$\Gamma$, the splitting is smaller than the experimental resolution for energies $E\leq 0.6\,\mathrm{meV}$, which imposes a conservative lower bound on the magnon velocity:
\begin{equation}
{v}_{\mathrm{R}\Gamma}> \frac{0.6\,\mathrm{meV}}{\frac{1}{2}\cdot f_{\left(\bar{H}\bar{H}H\right)}}\approx 30\cdot\frac{\mathrm{meV}}{\mathrm{r.l.u.}},
\end{equation}
where we used $1\,\text{r.l.u.}\approx 1.340\,\text{\AA}^{-1}$ for CeIn$_3$ at ambient pressure.

We assume a gapless magnon dispersion and determine the magnon velocities by linear weighted least-squares fits to the data points inferred from constant-energy cuts at energies $E=0.8\,\mathrm{meV}$, $1\,\mathrm{meV}$, .., $1.6\,\mathrm{meV}$. We performed four fits, where we considered the $N=2$, $3$, ..., $5$ data points with lowest energies and inferred the mean values. For the cubic face diagonal, the same energies $E=0.8\,\mathrm{meV}$, $1\,\mathrm{meV}$, .., $1.6\,\mathrm{meV}$ were considered and fits were carried out for the $N=2$, $3$, ..., $5$ data points with lowest energies. In turn, for the cubic edge, constant-energy cuts at $E=0.6\,\mathrm{meV}$, $0.8\,\mathrm{meV}$, .., $1.6\,\mathrm{meV}$ were considered and fits were carried out for the $N=2$, $3$, ..., $6$ data points with lowest energies.

\begin{table}
{
\caption{\textbf{Magnon velocities at the R point as inferred from experimental data. Shown are the values along the cubic space diagonal R$\Gamma$, the cubic face diagonal RX, and the cubic edge RM}. The magnon velocity was determined by least square fits considering several sets of sample points that are described in the text. Uncertainties correspond to statistical errors.}\label{tab:magnon velocities}}
\centering
\begin{tabular}{c|P{2cm}P{2cm}P{2cm}}
Number of sampling points:&\multicolumn{3}{c}{\textbf{Magnon velocities ${v}_{\Delta \bf{q}}$ ($\mathrm{meV}/ \, \mathrm{r.l.u.}$)}}\\
\centering
N&$\Delta\bf{q}\parallel$R$\Gamma$ & $\Delta\bf{q}\parallel$RX& $\Delta\bf{q}\parallel$RM\\\hline
\centering
2&42.2 $\pm$ 4.2 &35.5 $\pm$ 4.3&47.0 $\pm$ 0.1\\
3&38.6 $\pm$ 3.3 &37.3 $\pm$ 2.5&43.8 $\pm$ 3.8\\
4&36.7 $\pm$ 2.8 &37.4 $\pm$ 1.9&41.5 $\pm$ 3.2\\
5&34.6 $\pm$ 1.9 &36.1 $\pm$ 1.8&37.7 $\pm$ 3.4\\
6& & &36.5 $\pm$ 2.9\\\hline
mean$\pm$std& 38.0   $\pm$ 1.6  &36.6 $\pm$ 0.5&41.3  $\pm$ 1.9
\end{tabular}
\end{table}

The magnon velocities obtained from the fits as well as the mean values for the three reciprocal space directions are presented in Tab.~\ref{tab:magnon velocities}. The velocity averaged over the three directions R$\Gamma$, RX, and RM is given by ${v}_{\Delta \bf{q}}=(38.6\pm0.8)$ $\mathrm{meV}/ \, \mathrm{r.l.u.}$.

\subsection{Magnon velocities inferred from MO-PAM calculations}

The magnon dispersion inferred from MO-PAM calculations displays essentially linear behavior at energies between $0\,\mathrm{meV}$ and $1\,\mathrm{meV}$ for the three reciprocal-space directions R$\Gamma$, RX, and RM. The slopes for the three directions as inferred from weighted least squares fits with a linear function passing through the origin result in the velocities ${v}_{\mathrm{R}\Gamma}=\left(43.5\pm1.5\right)$ $\mathrm{meV}/ \, \mathrm{r.l.u.}$, ${v}_{\mathrm{RX}}=\left(45.2\pm1.7\right)$  $\mathrm{meV}/ \, \mathrm{r.l.u.}$, and ${v}_{\mathrm{RM}}=\left(47.4\pm1.2\right)$ $\mathrm{meV}/ \, \mathrm{r.l.u.}$. The velocity averaged over the three directions R$\Gamma$, RX, and RM is given by ${v}_{\Delta \bf{q}}=(45.3\pm0.9)$ $\mathrm{meV}/ \, \mathrm{r.l.u.}$.

\section{Bandwidth}

The bandwidth of magnetic excitations was determined on the dispersion on the subpath $\Gamma$XM$\Gamma$ (note, that the subpath R$\Gamma$ is not fully covered by INS data). The dispersion obtained via our experiment has the bandwidth $W_{\mathrm{exp}}=\left(2.75\pm0.03\right)\,\mathrm{meV}$, whereby the maximum of the recorded dispersion is attained on the subpath M$\mathrm{\Gamma}$. The dispersion inferred from MO-PAM calculations has a slightly larger bandwith given by $W_{\mathrm{MO-PAM}}=\left(3.18\pm0.02\right)\,\mathrm{meV}$ and the corresponding maximum is attained on the subpath $\Gamma$X.

The size of the bandwidth is well reproduced by fits with Heisenberg models. The bandwidths of the $J_1$ fit, $J_1$-$J_2$ fit, $J_1$-$J_2$-$J_3$ fit, and $J_1$-$J_2$-$J_3$-$J_4$ fit on the path R$\Gamma$XM$\Gamma$ are given by $2.91\,\mathrm{meV}$, $3.24\,\mathrm{meV}$, $3.02\,\mathrm{meV}$, and $3.11\,\mathrm{meV}$, respectively. Here $J_\nu$ denotes the Heisenberg exchange connecting the $\nu$th nearest neighbor pair.

\section{Ratio of magnon velocity and bandwidth}

The steep dispersion at the R point manifests itself in the extremely large ratio of magnon velocity to bandwidth, which is typically expressed in dimensionless units. The ratio as inferred from experimental data amounts to:
\begin{equation}
\eta_{\text{exp}}=\left(\frac{{v}_{\Delta \bf{q}}}{W}\right)_{\mathrm{exp}}\cdot \frac{1\cdot\mathrm{r.l.u.}}{2\pi}= 2.23\pm0.06 .
\end{equation}

Calculations that are based on MO-PAM result in a similarly large value given by:
\begin{align}
\eta_{\text{MO-PAM}}=\left(\frac{{v}_{\Delta \bf{q}}}{W}\right)_{\text{MO-PAM}}\cdot \frac{1\cdot\mathrm{r.l.u.}}{2\pi}=2.27\pm0.05.
\end{align}

In contrast, finite series expansions using exchange constants up to a few nearest-neighbors typically feature substantially smaller values. The corresponding fits are shown in section VII. The $J_1$ fit, $J_1$-$J_2$ fit, $J_1$-$J_2$-$J_3$ fit, and $J_1$-$J_2$-$J_3$-$J_4$ fit lead to $\eta=\frac{1}{\sqrt{3}}\approx0.58$, $\eta=0.56$, $\eta=0.90$, and $\eta=0.91$ (for the ratio of magnon velocity at R in direction R$\Gamma$ to bandwidth on the path R$\Gamma$XM$\Gamma$), respectively.

\section{Fit of experimental data by Heisenberg models}

The magnon dispersion on the path R$\Gamma$XM$\Gamma$, which is presented in Fig.~3 of the main text, was fitted with low-order nearest neighbour Heisenberg models. To reproduce the bandwidth of measured magnons, the models were fitted by means of least-squares statistics to the 14 data-points that were recorded in the high-energy setting.

The excitation dispersion of a Heisenberg model can be expressed in terms of:
\begin{align}
\label{supplement:equation:HeisenbergExpansion1}
E\left(\bf{q}\right)=\frac{1}{2} \sqrt{\left|\left(\gamma_{\bf{q}_{\mathrm{AFM}}}-\gamma_{\bf{q}}\right)\cdot \left(\gamma_{\bf{q}_{\mathrm{AFM}}}-\gamma_{\bf{q}_{\mathrm{AFM}}+\bf{q}}\right)\right|},
\end{align}
with $\bf{q}_{\mathrm{AFM}}=(\frac{1}{2},\frac{1}{2},\frac{1}{2})$ and  $\bf{q}=(q_1,q_2,q_3)$. Accounting for exchange constants up to fourth nearest neighbours, $\gamma_{\bf{q}}$ is given by:
\begin{align}
\label{supplement:equation:HeisenbergExpansionGamma}
\gamma_{\bf{q}}&=2 J_1\left[\cos(2\pi q_1)+\cos(2 \pi q_2)+\cos(2 \pi q_3)\right] \nonumber \\
&\quad +2 J_2\left[\cos(2\pi (q_1+q_2))+\cos(2 \pi (q_1-q_2)) +\cos(2 \pi (q_3+q_1)) \right. \nonumber\\ 
&\quad \quad \quad \left.+\cos(2 \pi (q_3-q_1))+\cos(2 \pi (q_2+q_3))+\cos(2 \pi (q_2-q_3))\right]\nonumber\\ 
&\quad +2 J_3\left[\cos(2 \pi (q_1+q_2+q_3))+\cos(2 \pi (-q_1+q_2+q_3))\right.\nonumber\\
&\quad \quad \quad \left. +\cos(2 \pi (q_1-q_2+q_3)) +\cos(2 \pi (q_1+q_2-q_3))\right] \nonumber\\
&\quad + 2 J_4 \left[  \cos(4 \pi q_1)+\cos(4 \pi q_2)+\cos(4 \pi q_3) \right].
\end{align}

\begin{figure*}
\includegraphics{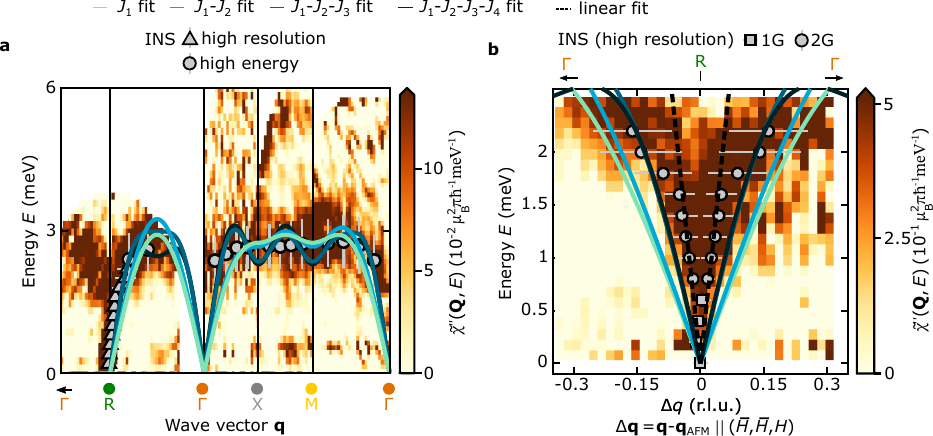}
\caption{\label{fig:S14}\textbf{Fit of high-energy experimental data by Heisenberg models.} The neutron scattering data obtained via the high-energy setting were fitted by a $J_1$ Heisenberg-model, $J_1$-$J_2$ model, $J_1$-$J_2$-$J_3$ model, and $J_1$-$J_2$-$J_3$-$J_4$ model, respectively, as explained in the text. \textbf{a} Comparison of these fits (solid lines) with data of the high-energy setting (color-coding). The triangle and circle symbols correspond to maxima of Gaussian fits to cuts through the high-resolution and high-energy INS data, respectively. Error bars denote the standard deviations of these profiles. \textbf{b} Comparison of the fits (solid lines) with data of the high-resolution setting (color-coding). The square and circle symbols correspond to maxima of single and double Gaussian profiles, respectively, fitted to cuts through the high-resolution INS data. Error bars denote the standard deviations of these profiles. The doted lines correspond to linear fits to the INS data.
}
\end{figure*}

The $J_1$-fit (for which we set $J_2=J_3=J_4=0$) resulted in $J_1=0.97$, the $J_1$-$J_2$-fit ($J_3=J_4=0$) in $J_1=1.34$ and $J_2=0.13$, the $J_1$-$J_2$-$J_3$-fit ($J_4=0$) in $J_1=0.38$, $J_2=0.04$, and $J_3=0.52$, and the $J_1$-$J_2$-$J_3$-$J_4$-fit in $J_1=0.22$, $J_2=0.00$, $J_3=0.69$, and $J_4=0.16$ (in units $\mathrm{meV}$). Figure~\ref{fig:S14} shows the four Heisenberg fits in comparison with experimental data. The bandwidth that was recorded by INS is reproduced by each of the fits. In contrast, the magnon-velocity at the R point, as inferred by a linear fit to INS data, is distinctively larger than the slope of each of the Heisenberg fits.

\section{Currat-Axe Spurions and Steep Magnon Dispersions}

We now illustrate why triple-axis spectroscopy is not well-suited to investigate the steep dispersion that we observe in CeIn$_3$. This is due to spurious scattering that is caused by neutrons scattered incoherently at the monochromator or analyzer crystals of a triple-axis spectrometer. When the spectrometer is set to a non-zero energy transfer, i.e., $\vert \bf{K}_i\vert \neq \vert \bf{K}_f\vert$, but the angle between $\bf{K}_i$ and $\bf{K}_f$ as well as the orientation of the sample correspond to a geometry that allows for Bragg scattering from the sample, the incoherently scattered neutrons from the monochromator that have a wave vector $\vert\bf{K}_{i,\text{inc}}\vert = \vert \bf{K}_f\vert$ lead to the observation of accidental Bragg scattering. These scattering events, which may also alternatively arise from neutrons incoherently scattered at the analyzer, result in  spurious intensity that appears as a dispersive feature close to a magnetic zone center with a linear dispersion. In the literature, this is well-known as Currat-Axe spurion~\cite{2002_Shirane_TAS}.

To show this, we have repeated our experiment on CeIn$_3$ previously carried out on CNCS at ORNL, using the multiplexing triple-axis spectrometer CAMEA~\cite{2020_Lass_CAMEA}. For each of the three incident neutron energies, spectroscopy data were recorded with two different positions of the multiplexing detector, namely $2\theta=-45\,\mathrm{deg}$ and $-41\,\mathrm{deg}$, and with different angles of vertical sample rotation covering an angular range of $60\,\mathrm{deg}$ with a step size of $0.5\,\mathrm{deg}$.

\begin{figure*}
\includegraphics{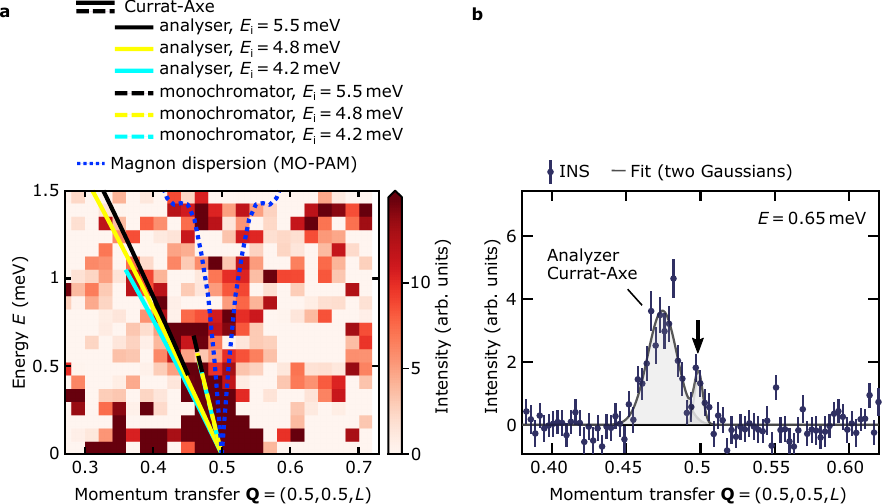}
\caption{\label{fig:Supplement_Currat-Axe}\textbf{Triple axis spectroscopy and Currat-Axe spurions} Shown in \textbf{a} is neutron spectroscopy intensity around momentum transfers $(0.5,0.5,0.5)$, which represents the R-point. Background recorded at 20 K were subtracted from the foreground data recorded at 2 K. INS intensity was integrated within a reciprocal space distance of $\pm0.09\,\mathrm{r.l.u.}$ along the axis $(1\bar{1}0)$. The location of Currat-Axe spurions from the analyzer crystal and from the monochromator are indicated by solid and interrupted lines, respectively. The spurions were calculated for incident neutron energies $4.2\,\mathrm{meV}$ (cyan color), $4.8\,\mathrm{meV}$ (yellow color), and $5.5\,\mathrm{meV}$ (black), respectively. The blue dotted line shows the trajectory of the magnon dispersion resulting from MO-PAM calculations. Shown in \textbf{b} is an exemplary $\bf{Q}$-cut at constant energy $0.65\,\mathrm{meV}$. INS intensity was integrated over a reciprocal space distance $\pm0.035\,\mathrm{r.l.u.}$ parallel to the axis $(1\bar{1}0)$ and over energies between $0.5\,\mathrm{meV}$ and $0.8\,\mathrm{meV}$. The error bars denote the statistical error. The binning for the $(001)$ direction amounts to $0.02\,\mathrm{r.l.u.}$. The data were fitted with a superposition of two Gaussian profiles. The peak at smaller $L$ arises from spurious scattering from the analyzer, whereas the peak in the center (black arrow) may be a superposition of spurious scattering from the monochromator and of magnons.}
\end{figure*}

The branches of Currat-Axe spurions originating from the monochromator as well as from the analyzer were calculated by means of the software \textit{MJOLNIR}~\cite{2020_Lass_SoftwareX} and are shown in Fig.~\ref{fig:Supplement_Currat-Axe}. Note that these lines disperse in close vicinity of the real magnon dispersion, that is, however, substantially weaker. In the constant-energy cut shown in \textbf{b} the Currat-Axe spurions from incoherent analyzer scattering result in a peak at around $L=0.47\,\mathrm{r.l.u.}$ (Gaussian at smaller $L$). The signal in the center (Gaussian at larger $L$ indicated by a black arrow) may be superposition of magnon scattering as well as Currat-Axe spurions arising from incoherent scattering at the monochromator.

\section{Previous experimental studies of magnetic excitations}
\begin{figure*}
\includegraphics{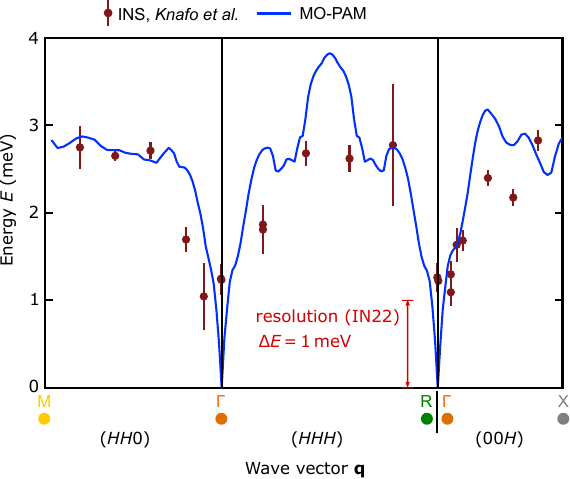}
\caption{\label{fig:S16}\textbf{Comparison of MO-PAM calculations with previous INS-data.} Experimental data that were measured at IN22 and reported by \textit{Knafo et al.} \cite{2003_Knafo_JPhysCondensMatter} are compared with MO-PAM calculations from this study. The red arrow indicates the size of the instrumental resolution.}
\end{figure*}

The magnetic excitations in CeIn$_3$ were previously characterised by \textit{Knafo et al.} \cite{2003_Knafo_JPhysCondensMatter} using triple-axis spectroscopy. The data, which are presented in Fig.~\ref{fig:S16}, are qualitatively in agreement with our study in major parts of the Brillouin zone. In particular, the recorded dispersion features the same bandwidth that we identified in our study. In the vicinity of $\Gamma$ and in the vicinity of R, the energy resolution of IN22 was not sufficient to resolve the extremely steep excitation dispersion.

The authors of Ref.~\cite{2003_Knafo_JPhysCondensMatter} fitted the dispersion of excitations with a $J_1$-$J_2$ model, which we identify in our report as inadequate to account for the excitations in CeIn$_3$ (see Sec.~VII). In particular, the $J_1$-$J_2$ model is unable to reproduce the relatively large ratio of magnon velocity to bandwith, which we observed. In addition, the authors of Ref.~\cite{2003_Knafo_JPhysCondensMatter} inferred a gap of size 1.28 meV at the magnetic zone center, i.e., at the R point, which is in stark contrast to our results indicating a gapless dispersion at the magnetic zone center.

\section{Comparison of DFT calculations with the band structure obtained by photo emission studies}\label{sec:ARPES}

\begin{figure*}
\includegraphics{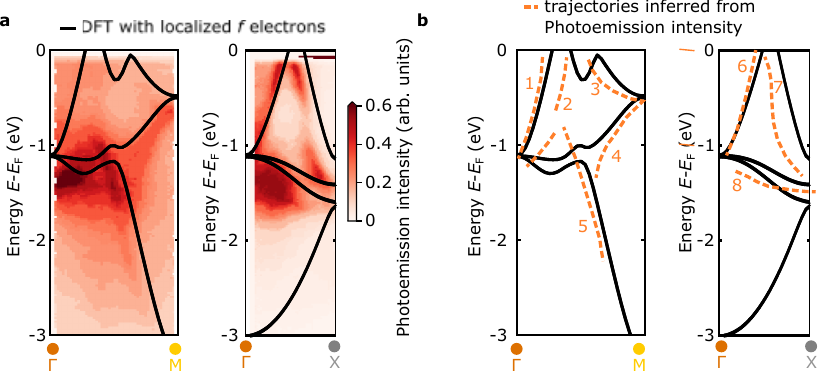}
\caption{\label{fig:S17}\textbf{Comparison of DFT-calculations with angle-resolved photo emsission spectroscopy (ARPES) studies.} \textbf{a} Polarization dependent valence band structure obtained by ARPES measurements, as reported in Ref.~\cite{2016_Zhang_SciRepa} and provided in terms of the color-coding, as compared to the band structure from our DFT calculations obtained for localised $f$-electrons (black lines). The ARPES data were extracted from Fig.~2 of Ref.~\cite{2016_Zhang_SciRepa}. \textbf{b} Trajectories of the band structure inferred from local maxima of the ARPES data (orange dotted lines) compared with the trajectories obtained from our DFT calculations (black lines).}
\end{figure*}

Fig.~\ref{fig:S17} presents a comparison of the DFT band structure shown in Fig.~\ref{fig:DFT_TB_comparison_nof} with the polarization dependent valence band structure obtained by angle-resolved photo emission spectroscopy (ARPES) measurements (cf. Ref.~\cite{2016_Zhang_SciRepa}). Considering the challenges that arise when comparing surface sensitive ARPES data to DFT calculations as well as difficulties in ARPES experiments on cubic materials, the agreement between DFT and ARPES is excellent. Notably, in cubic materials the momentum resolution along the surface normal can be broadened resulting in difficulties of interpreting the ARPES data \cite{2022_Rahn}.

\newpage

\providecommand{\noopsort}[1]{}\providecommand{\singleletter}[1]{#1}%
%
